\documentclass[12pt,letterpaper,nohyper]{JHEP3}

\usepackage{amsmath,amssymb,amsfonts,theorem,epsfig,slashed,xspace}

\title{Quantum Attractor Flows}

\preprint{LPTENS-07-26}

\author{M.~G\"unaydin$^{a,b}$, A.~Neitzke,$^b$, B. Pioline$^{c,d}$ and 
A.~Waldron$^e$\footnote{Email: {\tt murat@phys.psu.edu}, 
{\tt neitzke@post.harvard.edu}, {\tt pioline@lpthe.jussieu.fr}, 
{\tt wally@math.ucdavis.edu}}\\

$^a$ Physics Department, Pennsylvania State University,\\
University Park, PA 16802, USA\\

$^b$ School of Natural Sciences, Institute for Advanced Study, Princeton, 
NJ, USA\\

$^c$ Laboratoire de Physique Th\'eorique et Hautes 
Energies\footnote{Unit\'e mixte de recherche du CNRS UMR 7589},~
Universit\'e Pierre et Marie Curie - Paris 6,
4 place Jussieu, F-75252 Paris cedex 05 \\

$^d$ Laboratoire de Physique Th\'eorique de l'Ecole Normale
Sup\'erieure\footnote{Unit\'e mixte
de recherche du CNRS UMR 8549}~,\\
24 rue Lhomond, F-75231 Paris cedex 05\\

$^e$ Department of Mathematics, One Shields Avenue,\\
University of California, Davis, CA 95616, USA}

\abstract {
Motivated by the interpretation of the Ooguri-Strominger-Vafa
conjecture as a holographic correspondence in the mini-superspace
approximation, we study the radial quantization of stationary,
spherically symmetric black holes in four dimensions. A key ingredient
is the classical equivalence between the radial evolution equation
and geodesic motion of a fiducial particle on the 
moduli space $\cM^*_3$ of the three-dimensional theory
after reduction along the time direction.
In the case of $\cN=2$ supergravity, $\cM^*_3$ is a 
para-quaternionic-K\"ahler manifold; in this case, we show that BPS
black holes correspond to a particular class of geodesics which lift
holomorphically to the twistor space $\cZ$ of $\cM^*_3$,
and identify $\cZ$ as the BPS phase space. We give a natural
quantization of the BPS phase space in terms of the sheaf cohomology of
$\cZ$, and compute the exact wave function of a BPS black hole with
fixed electric and magnetic charges in this framework. 
We comment on the relation to the topological string amplitude, 
extensions to $\cN>2$ supergravity theories, and applications
to automorphic black hole partition functions.}

\newcommand{\ophi}{\stackrel{\circ}{\phi}}
\renewcommand{\Im}{{\rm Im}}
\renewcommand{\Re}{{\rm Re}}
\newcommand{\pa}{\partial}
\newcommand{\CN}{{\cal N}}

\newcommand{\nn}{\nonumber}
\newcommand{\IR}{\mathbb{R}}
\newcommand{\IC}{\mathbb{C}}
\newcommand{\IZ}{\mathbb{Z}}

\newcommand{\Tr}{\mbox{Tr}}

\newcommand{\half}{\frac12}
\newcommand{\tzeta}{\tilde\zeta}
\newcommand{\zetat}{\tilde\zeta}
\newcommand{\eps}{\epsilon}

\newcommand{\txi}{{\tilde\xi}}
\newcommand{\ri}{{\rm i}}
\newcommand{\rj}{{\rm j}}
\newcommand{\rk}{{\rm k}}
\newcommand{\ra}{{\rm a}}
\newcommand{\rb}{{\rm b}}

\newcommand{\cC}{\mathcal{C}}

\newcommand{\cK}{\mathcal{K}}
\newcommand{\cM}{\mathcal{M}}
\newcommand{\cN}{\mathcal{N}}

\newcommand{\cS}{\mathcal{S}}

\newcommand{\cV}{\mathcal{V}}
\newcommand{\cJ}{\mathcal{J}}
\newcommand{\N}{{\mathcal N}}

\newcommand{\hk}{{hyperk\"ahler}\xspace}
\newcommand{\qk}{{quaternionic-K\"ahler}\xspace}
\newcommand{\C}{{\mathbb C}}
\newcommand{\PP}{{\mathbb P}}
\newcommand{\R}{{\mathbb R}}
\newcommand{\cZ}{\mathcal{Z}}
\newcommand{\cO}{\mathcal{O}}
\newcommand{\abs}[1]{\lvert#1\rvert}

\newcommand{\inprod}[1]{\langle#1\rangle}

\def\bea{\begin{eqnarray}}
\def\eea{\end{eqnarray}}
\def\be{\begin{equation}}
\def\ee{\end{equation}}
\def\ba{\begin{align}}
\def\ea{\end{align}}
\def\bse{\begin{subequations}}
\def\ese{\end{subequations}}

\theoremstyle{plain}

\begin{document}


\numberwithin{equation}{section}

\section{Introduction}
In view of the inherent difficulties in quantizing Einstein's gravity, 
mini-superspace models of quantum gravity, where all but a finite number 
of degrees of freedom consistent with certain symmetries are retained, 
have been a popular subject of study, particularly in quantum cosmology. 
A space-like analogue of these cosmological models, the radial quantization of
static, spherically symmetric black holes in  Einstein and Einstein-Maxwell 
gravity has also been much studied~\cite{Thiemann:1992jj,Kuchar:1994zk, 
Cavaglia:1994yc,Hollmann:1996cb,Hollmann:1996ra,Breitenlohner:1998yt}.
In the present work we instill supersymmetry in these early
treatments and lay out a quantization scheme for stationary, 
spherically symmetric solutions of 
four-dimensional $\CN=2$ supergravity. Our motivation 
stems from recent developments in black hole and string physics,
which we now briefly review.

\subsection{Motivation}
The microscopic origin  of the geometric entropy of 
supersymmetric black holes in type~IIA string theory compactified on 
a Calabi-Yau three-fold $Y$ 
can be investigated by virtue of
several simplifying properties:
\begin{itemize}
\item[(i)] The ``attractor phenomenon''~\cite{Ferrara:1995ih,
Ferrara:1996um,Ferrara:1997tw} implies that
the entropy and scalar fields at the horizon (hence the K\"ahler moduli 
of $Y$) are functions of the electric and magnetic charges only.
\item[(ii)] Since BPS black holes are extremal, they  are not subject to
Hawking evaporation, yet their entropy can be made as large as
desired by increasing their charges.
\item[(iii)] Being supersymmetric, they are expected to correspond to
exact zero-energy eigen-states (or eigen-matrices) 
of the microscopic Hamiltonian.
\item[(iv)] Due to the tree-level decoupling between vector multiplets and
hypermultiplets, the string coupling may be made
as small as desired, such that  micro-states
can be described as a gas of weakly coupled open strings or membranes, 
whose microscopic entropy can be reliably computed
on combinatorial grounds. 
\end{itemize}
Taken together, these simplifications have led to a clear microscopic 
derivation of the Bekenstein-Hawking entropy of a class of BPS 
black holes~\cite{Strominger:1996sh,Maldacena:1996gb,Johnson:1996ga}, 
accurate in the limit of large charges (even
reproducing the first subleading correction in the M-theory 
approach~\cite{Maldacena:1997de}).
The modern version of this argument uses
holographic duality between M-theory on the attractor near-horizon 
geometry $[AdS_3/\Gamma] \times S^2\times Y_*$ of a five-dimensional
black string whose reduction to four dimensions produces the 
black hole of interest, and a two-dimensional superconformal field 
theory at the boundary of $AdS_3$ (see {\it e.g.}~\cite{Kraus:2006wn,
Pioline:2006ni} for reviews and references).

Recently, there have been many efforts to extend this agreement 
beyond the large charge regime. On the macroscopic side the 
geometric entropy, including the effects of an infinite series 
of higher-derivative BPS couplings in the low energy effective action,
has been computed~\cite{LopesCardoso:1999cv,LopesCardoso:1999ur,
LopesCardoso:1999xn}; the result takes a particularly simple form when 
expressed in terms of a mixed thermodynamical ensemble with fixed magnetic
charges $p^I$ and electric potentials $\phi^I$~\cite{Ooguri:2004zv}. 
Combined with the relation between
higher-derivative BPS couplings and the topological string amplitude
on the Calabi-Yau threefold $Y$, it suggests a intriguing relation
\cite{Ooguri:2004zv}
\be
\label{osvwig}
\Omega(p^I, q_I) \sim \int d\phi^I \
\Psi_{\rm top}^*( p^I - i \phi^I)\  
\Psi_{\rm top}( p^I + i \phi^I)\  
e^{\pi \phi^I q_I} \, ,
\ee
between the indexed degeneracies $\Omega(p^I, q_I)$ of BPS states 
with magnetic and electric 
charges $(p^I,q_I)$, and the topological string amplitude $\Psi_{\rm top}$; 
the latter should be understood as a wave function in the 
real (background-independent) polarization ensuring covariance under
a change of electric-magnetic 
duality frame~\cite{Witten:1993ed,Gerasimov:2004yx,Verlinde:2004ck,
Gunaydin:2006bz}. The equality in
\eqref{osvwig} was conjectured to hold to all orders in an expansion at 
large charges, as supported by various explicit checks 
for compact~\cite{Dabholkar:2005dt,Dabholkar:2005by} 
and non-compact~$Y$~\cite{Vafa:2004qa,Aganagic:2004js,Dijkgraaf:2005bp}. 
The relation~\eqref{osvwig} 
has been derived recently by evaluating the
elliptic genus of M-theory in the above near-horizon geometry 
\cite{Gaiotto:2006ns,Gaiotto:2006wm, deBoer:2006vg,Kraus:2006nb} 
(see also~\cite{Denef:2007vg} for an alternative approach using D6-branes). 

Both these recent discussions of the subleading corrections to the entropy,
as well as the original derivations in~\cite{Maldacena:1996gb,
Johnson:1996ga,Maldacena:1997de}, rely on the possibility
of lifting the four-dimensional black hole to a five-dimensional black string: 
while this is indeed possible for vanishing or unit $D6$-brane charge, 
in general
the five-dimensional parent is a black hole in a singular Taub-NUT
background, possibly accompanied by a black 
ring~\cite{Gaiotto:2005gf,Gaiotto:2005xt}. 
In fact, standard holography arguments suggest that it should 
be possible to describe the spectrum of black hole micro-states
in terms of superconformal quantum mechanics on the (disconnected) 
boundary of the near-horizon geometry $AdS_2\times S^2\times Y_*$. 
Unfortunately, this superconformal quantum mechanics has remained
vexingly elusive (see however~\cite{Gaiotto:2004ij,Gaiotto:2005rp} for 
some recent progress).

\begin{figure}
\centerline{\hfill
\includegraphics[height=3cm]{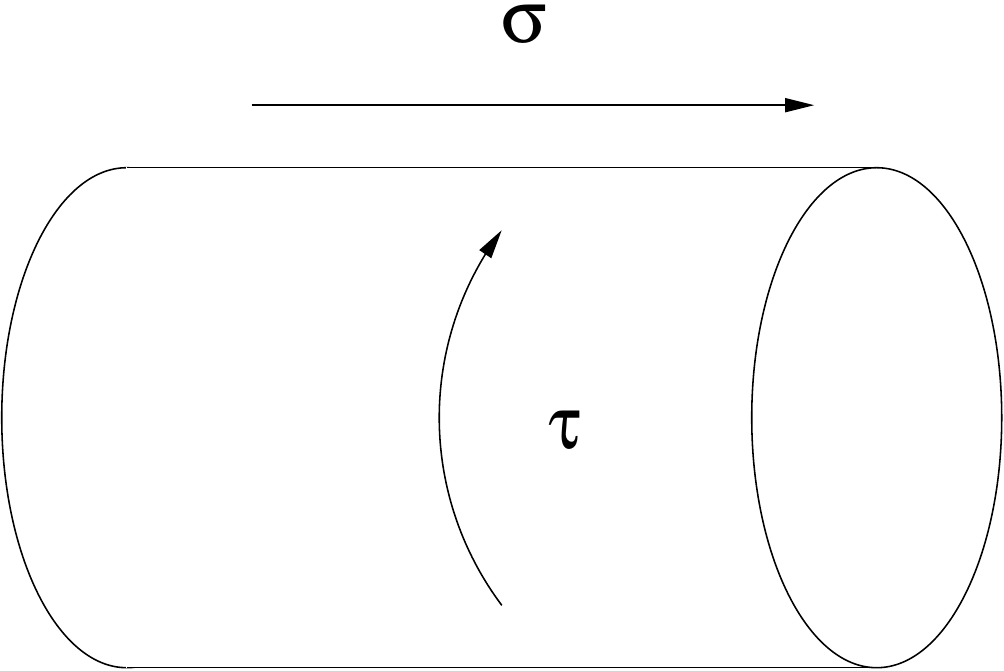}\hfill
\includegraphics[height=5cm]{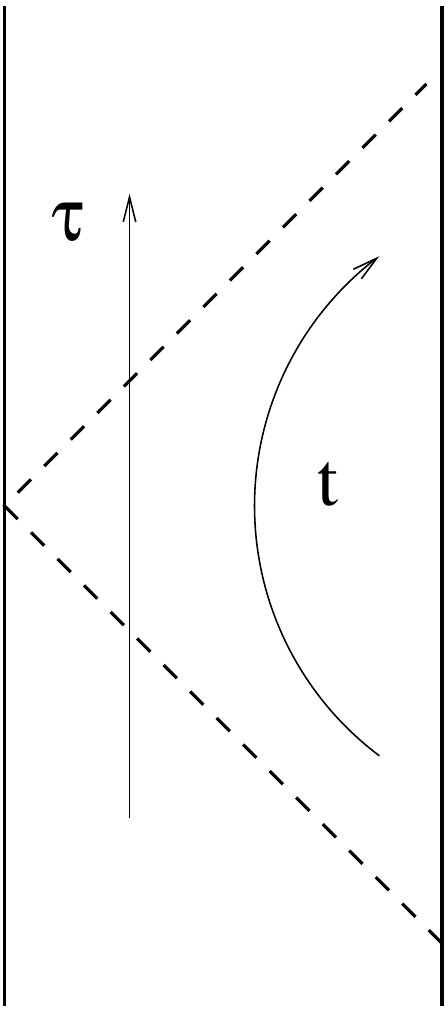}
\hfill}
\caption{{\it Left:} the cylinder amplitude in string theory
can be viewed either as 
a trace over the open string Hilbert space (quantizing along $\tau$)
channel) or as an inner product between two wave functions in the
closed string Hilbert space (quantizing along $\sigma$). 
{\it Right:}
The global geometry of Lorentzian $AdS_2$ has the topology of a strip; 
its Euclidean continuation at finite temperature becomes a cylinder.
$\tau$ and $t$ are the global and Poincar\'e time, respectively.
\label{cylpic}}
\end{figure}

Lacking a concrete definition of the superconformal quantum mechanics
on the boundary of $AdS_2$, it is worthwhile trying to obtain 
indirect information on its spectrum using the AdS/CFT correspondence.
Specifically, the cylinder-like topology of thermal $AdS_2$ 
suggests, in analogy with the familiar open/closed string duality,
that it should be possible to derive the partition function of the
black hole micro-states -- the ``open string channel'' -- as an overlap
of wave functions in a radial quantization scheme -- the ``closed string
channel'' -- see Figure \ref{cylpic}. 
Performing a radial quantization of gravity is hardly doable 
in general, but becomes tractable in a ``mini-superspace'' truncation 
where only stationary spherically symmetric geometries are retained.

It has been proposed to 
interpret~\eqref{osvwig} in just this way~\cite{Ooguri:2005vr}: 
regard the left-hand side as the
partition function in the Hilbert space of BPS black holes 
with given values
of the charges (and zero Hamiltonian), and the right-hand side as 
the overlap of two wave functions in the Hilbert 
space\footnote{It should be stressed that, just as in conformal field 
theory on the cylinder, there is no relation between the
spectrum in the open and closed string channels, until 
string interactions are introduced.}
of spherically symmetric BPS geometries. To spell this out, analytically 
continue $\phi^I=i\chi^I$ to the imaginary axis and define
\be
\label{osvwig2}
\Psi_{p,q}^\pm(\chi) \equiv 
e^{\pm \frac{i \pi \chi q }{2}}\ \Psi_{\rm top}(\chi\mp p)
\equiv V_{p,q}^{\pm} \cdot  \Psi_{\rm top}(\chi)\ .
\ee
Equation~\eqref{osvwig} may then be rewritten more suggestively as 
an overlap of two wave functions,
\be
\label{osvlap}
\Omega(p,q) \sim \int d\chi\ [ \Psi^-_{p,q}(\chi)]^* \ \Psi_{p,q}^+(\chi)\ .
\ee
This interpretation assumes one can view
the topological amplitude $\Psi_{\rm top}$
as a wave function for the radial quantization of spherically symmetric
geometries;
if true, it would provide a physical interpretation for
the wave function property of the topological
string partition function, observed at a formal level 
in~\cite{Witten:1993ed}. 

While the mini-superspace approximation is usually at best ill-controlled, 
one may hope that,
for the purpose of the indexed partition function of BPS black holes, the
truncation to BPS ground states in the radial channel may be justified.
In this respect, note that the quantization of BPS configurations 
has been applied in various set-ups~\cite{Mandal:2005wv,
Maoz:2005nk,Grant:2005qc,Biswas:2006tj,Mandal:2006tk}, and used for 
a derivation of the entropy of two-charge
black holes~\cite{Rychkov:2005ji}.

Finally, we note that further interest in the 
quantization of attractor flows arises from the 
analogy between black hole attractor equations and the
equations that determine supersymmetric vacua in flux 
compactifications, and possible applications of the black hole
wave function to vacuum selection in string theory~\cite{Ooguri:2005vr}. 

\subsection{Summary and Outline}

Some of our results have been announced 
in~\cite{Gunaydin:2005mx,Pioline:2006ni}:
the key observation, explained in Section 2, 
is the equivalence~\cite{Breitenlohner:1987dg}  
between the radial equations of motion
for stationary, spherically symmetric solutions, and the geodesic motion
of a fiducial superparticle on a pseudo-Riemannian manifold ${\cal M}_3^*$;
the latter arises by supplementing the 
four-dimensional moduli space ${\cal M}_4$ with the various 
scalars arising in the dimensional reduction along the time direction. 
The electric and magnetic charges $(q_I,p^I)$
of the black hole, the ADM mass $m$ 
as well as the NUT charge\footnote{Bona
fide 4D black holes are obtained only for $k=0$, but keeping $k\neq 0$ is 
a key technical device.} $k$
are conserved Noether charges associated to isometries of 
${\cal M}_3^*$, whose Poisson brackets obey an extended Heisenberg algebra
\eqref{heis}. Extremal black holes 
correspond to light-like geodesics on ${\cal M}_3^*$. 
The phase space of stationary, 
spherically symmetric solutions is the cotangent bundle $T^*({\cal M}_3^*)$, 
or one of its symplectic quotients when some conserved charges are held fixed.
Quantization is then in principle clear:
the Hilbert space for radial quantization is the space of square-integrable
functions on ${\cal M}_3^*$, subject to the Hamiltonian and charge constraints.
In Subsection \ref{sphgeoq}, we briefly outline how physical observables can be
extracted from a wave function in this Hilbert space; we note however that 
conserved charges alone do not select a unique wave function.

This situation is vastly improved when restricting to BPS solutions.
As we show in Section \ref{sqm}, supersymmetry strongly restricts 
the allowed momentum along the geodesic,
effectively removing half
the degrees of freedom. In the context of $\CN=2$ supergravity, 
${\cal M}_3^*$ is an analytic continuation\footnote{\label{para-fn}
When discussing the case of $\cN=2$ supergravity in Section \ref{sqm}, 
we use for convenience the language of the Riemannian space ${\cal M}_3$, 
and complexify the coordinates.  All of our arguments and results could be
formulated in terms of the intrinsic para-quaternionic geometry of
$\cM_3^*$, which is the real slice directly related to the physical
problem at hand.} of a quaternionic-K\"ahler space ${\cal M}_3$
obtained by the $c$-map construction from the four-dimensional special
K\"ahler moduli space ${\cal M}_4$.  Supersymmetry requires the
momentum to satisfy certain quadratic constraints \eqref{susycond2}
built from the quaternionic vielbein
of  ${\cal M}_3$. The geometric structure of the BPS phase space
is however obscure in this formulation. 

Instead the supersymmetry constraint is better expressed 
by introducing the twistor space $\cZ$ -- 
a two-sphere bundle over ${\cal M}_3$ which carries a canonical 
complex
structure, as well as a K\"ahler-Einstein metric.
It is also useful to introduce the Swann bundle $\cS$ over ${\cal M}_3$, 
which is a line bundle over $\cZ$  with a hyperk\"ahler, 
$SU(2)$ and scale invariant metric.
Physically, the $S^2$ or $\IR^4$  fiber of $\cZ$ or $\cS$, respectively,
over ${\cal M}_3$ keeps track of the Killing 
spinor preserved by the black hole. Supersymmetric trajectories
on the base $\cM_3$ then simply correspond to ``holomorphic'' geodesics in one 
complex structure on $\cS$
({\it i.e.} trajectories whose tangent vector is holomorphic 
at any point), 
with no angular momentum in the fiber. These BPS geodesics
descend to holomorphic geodesics on $\cZ$.
The BPS phase space is then 
the twistor space $\cZ$ itself equipped with its K\"ahler form.
Thus, it is roughly
twice as small as the non-BPS phase space $T^*({\cal M}_3^*)$.

With this reformulation at hand quantization is 
again in principle clear: the BPS Hilbert space should
be the K\"ahler quantization of the twistor space $\cZ$.
Technically, this is complicated by the fact that $\cZ$
does not admit non-trivial holomorphic functions, and moreover 
has indefinite signature, due to the negative curvature of
the base $\cM_3$. Moreover, it should be possible to view
the BPS Hilbert space as a subspace of the unconstrained Hilbert 
space $L^2(G/K)$, determined by generalized harmonicity 
constraints \eqref{wavebpsqk} quantizing the classical quadratic
constraints \eqref{susycond2}.

There is a natural conjecture that addresses these concerns
all at once: the BPS Hilbert space should be 
the sheaf cohomology group $H^1(\cZ,{\cal O}(-\ell))$
for appropriate $\ell$. Indeed, there exists a generalized
Penrose transform which relates classes~in 
$H^1(\cZ,$ ${\cal O}(-\ell))$ to functions on the base $\cM_3$ 
solving exactly these
partial differential equations~\cite{MR664330,MR1165872,Neitzke:2007ke}. 
A special
case of this is the standard Penrose transform, which 
relates a cohomology class on a subset of $\C\PP^3$ 
to a solution of the conformal Laplacian
on a subset of $S^4$~(see \cite{Penrose:1972ia,ward-wells}). 
The value of $\ell$ determines the spin of the wave
function on $\cM_3$, and could in principle be computed by a 
careful quantization of the fermions in the one-dimensional
non-linear sigma model, which we defer to a forthcoming
publication~\cite{w-toappear}.

In contrast to the 
non-BPS case, specifying the conserved charges $p^I,q_I$ (at
vanishing NUT charge $k=0$) now
determines a unique wave function $\Psi_{p,q}$, a 
plane wave in the complex coordinates on $\cZ$ adapted 
to the Heisenberg symmetries. 
Contour integration of the BPS wave function on $\cZ$ leads to
the exact wave function~\eqref{exactbpswf} of a BPS black hole with charges
$(p,q)$ as a function of the four-dimensional vector-multiplet moduli,
as well as of the scale $U$ of the time direction\footnote{This
wave function was first computed in \cite{Neitzke:2007ke}, where
mathematical aspects of  the twistorial approach to black
holes were studied. In this paper we focus on the physical aspects of
this approach.}. The norm of the
wave function is maximal at the classical attractor point(s), but
is not exponentially suppressed away from them, contrary perhaps  
to expectations. In fact, the effective Planck constant grows 
as $e^{-U}$ toward the horizon at $U\to -\infty$, leading to large
quantum fluctuations. The implications of this result deserve to
be further investigated.

The outline of this paper is as follows. In Section 2 we review the
general equivalence between the radial evolution equations for stationary,
spherically black holes in 4 dimensions and geodesic flow on the
three-dimensional moduli space, and discuss the general features of the
radial quantization for non-supersymmetric black holes. In Section 3
we specialize to $\cN=2$ supergravity, show that 
twistor techniques allow one to characterize the geodesics associated
to BPS black holes, propose a natural quantization scheme 
of the BPS phase space, based on the K\"ahler quantization of the
twistor space, and compute the exact wave function for a BPS black hole 
in this framework (some of the material in this Section is a review
of the results in \cite{Neitzke:2007ke}). We conclude in Section 4 with
a discussion of the relation of our wave function to the topological 
string amplitude, 
applications to symmetric $\cN=2$ and $\cN>2$ supergravities,
and to automorphic counting functions for black hole micro-states,
and other directions.
In Appendix A we supply details on the reduction of the supersymmetry
conditions from 4 to 3 dimensions. In Appendix B we discuss
pure $\cN=2$ supergravity in 
four dimensions. In Appendix C, we comment on 
supergravity theories with $\cN=4$ and $\cN=8$. 

\section{Attractor Flows and Geodesic Motion}
We begin by reformulating the equations of motion for 
stationary solutions in four dimensions in terms of a gravity-coupled 
non-linear sigma-model on an extended moduli space ${\cal M}_3^*$ 
in three Euclidean 
dimensions. By assuming spherical symmetry the problem is
further reduced in Subsection \ref{sphgeo} to the geodesic motion 
of a fiducial particle on 
${\cal M}_3^*$. In Subsection \ref{sphgeoq} we quantize
this mechanical system. No assumption about
supersymmetry is made in this Section.

\subsection{Stationary Metrics and Harmonic Maps\label{statharm}}
We consider Einstein gravity in four dimensions coupled to $n_A$ Abelian
gauge fields $A_4^I$ and $n_S$ scalar fields $z^i$ with action
\be
\label{4dact}
S_4 =\! \int \!d^4 x \left[ -\frac12\sqrt{-\gamma} \ 
R[\gamma]  - g_{ij}\ dz^i \!\wedge {}^\star\! 
dz^{j} 
+ F^I \!\wedge\!\Big( 
\frac14 (\Im{\cal N})_{IJ}\wedge {}^\star\! F^J
\!-\frac18(\Re{\cal N})_{IJ} F^J \Big) \right] .
\ee
Here $\gamma$ denotes the four-dimensional metric, $g_{ij}$ ($i=1\dots n_S$)
the metric on the moduli space ${\cal M}_4$ where the (real) scalars $z^i$ take
their values, $F^I=dA_4^I$ ($I=1\dots n_A$) are the field strengths
of the Maxwell fields with complexified gauge couplings 
${\cal N}_{IJ}(z^i)=({\cal N}^{IJ})^{-1}$.

Now we restrict our attention to stationary configurations.
The most general stationary metric has the form
\be
\label{stationan}
ds_4^2 = -e^{2U} (dt + \omega)^2  + e^{-2U} ds^2_3  \ ,\quad
\ee
where the scalar $U$, one-form $\omega$ and line element $ds_3^2$ are 
functions on the spatial slice $\Sigma$ and independent of $t$.
Similarly we decompose the vector fields as
\be
A_4^I = \zeta^I (dt+\omega) + A^I_3\ ,
\ee
into pseudo-scalars $\zeta^I$ and one-forms $A^I_3$ defined on $\Sigma$
and assume that the scalars $z^i$ are independent of time.  The equations 
of motion for $(U, \omega, ds^2_3, \zeta^I, A^I_3)$ may be obtained
by reducing action $S_4$ along the time direction. In three dimensions,
the one-forms $A^I_\ri$ and $\omega$ can be dualized into 
axionic scalars $\tzeta_I$ and $\sigma$.
Thus, the four-dimensional theory reduces 
to a non-linear sigma model coupled to Euclidean gravity,
\be
\label{3dsig}
S_3 = \int\ \ d^3 x \left( \sqrt{g_3}\ R[g_3]
- g_{mn} \ d\phi^m \wedge \star d\phi^n \right) \ ,
\ee
whose the coordinates $\phi^m$ on the 
target space ${\cal M}_3^*$ include the
scalar fields $z^i$ from four dimensions together with $U$,  
$\zeta^I$, $\tzeta_I,\sigma$. In contrast to
the usual Kaluza-Klein reduction along a space-like direction,
the metric  $g_{mn}$ on ${\cal M}_3^*$ has indefinite signature:
\bea
\label{noncmap}
ds^2_{\cM_3^*} &=& 
dU^2+\frac12 g_{ij}dz^i dz^j
+e^{-4U} 
\left(d\sigma-\tilde \zeta_I d\zeta^I+\zeta^I d\tilde \zeta_I\right)^2 \\
 &&\hspace*{-15mm} + \frac12 e^{-2U} 
\left[  (\Im{\cal N})_{IJ} d\zeta^I d\zeta^J
+\, (\Im{\cal N})^{IJ}
\left(d\tilde\zeta_I+ (\Re{\cal N})_{IK} d\zeta^K\right)
\left(d\tilde\zeta_J+ (\Re{\cal N})_{JL} d\zeta^L\right) \right]\ ,\nn
\eea
(recall that $(\Im{\cal N})_{IJ}$ is negative definite).
It is related to its Riemannian counterpart ${\cal M}_3$ 
(from standard Kaluza-Klein reduction, see {\it e.g.}~\cite{Ferrara:1989ik})  
by analytic continuation $(\zeta^I,\tzeta_I)
\to  i(\zeta^I,\tzeta_I)$~\cite{Gunaydin:2005mx}. Thus, 
stationary solutions in four dimensions
are given by harmonic maps from the (in general curved) three-dimensional 
spatial slice to ${\cal M}_3^{*}$ \cite{Breitenlohner:1987dg}. 

Importantly, ${\cal M}_3^{*}$ 
possesses $2n+2$ isometries, reflecting symmetries of the stationary 
sector of the four-dimensional theory:  these are the
shift symmetries of $A^I, \tilde A_I,\omega$, as well as rescalings of 
time $t$. The Killing vector fields generating these isometries are
\be
\label{killingh}
p^I = \pa_{\tzeta_I} + \zeta^I \pa_\sigma\ ,\quad
q_I = \pa_{\zeta^I} - \tzeta_I \pa_\sigma\ ,\quad
k = \pa_\sigma\ ,\quad
m=-\pa_U - \zeta^I\pa_{\zeta^I} - \tzeta^I\pa_{\tzeta^I} - 2\sigma\pa_\sigma\ ,
\ee
and satisfy the Lie algebra
\be
\label{heis}
[ p^I, q_J ] = -2\delta^I_J\ k\ ,\quad
[m, p^I] = p^I\ ,\quad \ [m,q_I] = q_I\ ,\quad [m,k]=2k\ .
\ee
This notation anticipates the fact that the 
associated conserved quantities will be the
electric and magnetic charges, NUT charge, and ADM mass of
the black hole. In particular, the 
electric and magnetic charges $p^I,q_I$ satisfy an
Heisenberg algebra graded by the ADM mass $m$, with center $k$.

\subsection{Stationary, Spherically Symmetric Black Holes and Geodesics
\label{sphgeo}}
We now further restrict to spherically symmetric solutions.
The metric on the spatial slice $\Sigma$ can be parameterized as
\be
\label{3dsli}
ds_3^2 = N^2 (\rho)\ d\rho^2 + r^2(\rho) \ 
(d\theta^2 + \sin^2\theta\ d\phi^2)\ ,
\ee
while the scalars become functions of $\rho$ only.
The scalar curvature of $\Sigma$ is
\be
\sqrt{g_3}\ R^{(3)} = 
2 \sin\theta \left[ \frac{(r')^2}{N} + N - \frac{d}{d\rho} \left(
\frac{2 r r'}{N} \right) \right]\ ,
\ee
where the prime denotes a $\rho$--derivative.
Substituting in~\eqref{3dsig}, integrating over the angles $\theta,\phi$
and dropping  a total derivative term leads to
\be
\label{clasme}
S_1 = \int d\rho\ \left[ \frac{N}{2} + \frac{1}{2N} \left( 
r'^{2} - r^2 \, g_{mn}\,  \phi'^{m}\,  \phi'^{n} \right)  \right] \ .
\ee
This Lagrangian describes the motion of a fiducial particle on a 
cone\footnote{A
similar system arises in mini-superspace cosmology~\cite{Damour:2002et,
Pioline:2002qz}. Higher-derivative corrections to the geodesic motion
arising from $R^2$ corrections to the four-dimensional action have been 
discussed in \cite{Michel:2007vh}}
${\cal C}$ over the $d=3$ moduli space ${\cal M}_3^*$.
The  einbein $N$ on the particle worldline ensures invariance
under reparametrizations; its equation of motion enforces the mass shell 
condition
\be
\label{mscond}
r'^{2} - r^2 \, g_{mn}\, \phi'^{m} \, \phi'^{n} = N^2 \ ,
\ee
or equivalently, the Wheeler-De Witt (or Hamiltonian) constraint
\be
\label{hwdw}
H_{WDW} = (p_r)^2 - \frac{1}{r^2} g^{mn} p_m p_n - 1 \equiv 0 \ ,
\ee
where $p_r,p_m$ are the canonical momenta conjugate to $r$ and $\phi^m$.

Solutions are thus massive
geodesics on the cone $\cC$, with fixed unit mass. 
The motion separates into geodesic motion on the base of the cone
${\cal M}_3^*$, with 
affine parameter $\tau$ such that $d\tau=N\,d\rho/r^2$,
and motion along the radial direction $r$, 
\be
\label{rrhoc}
(p_r)^2 - \frac{C^2}{r^2} - 1 \equiv 0\ ,\quad  
g^{ab} p_a p_b \equiv C^2\ ,
\ee
where $p_r=r'/N$ and $p_m=r^2 \phi^{'m}/N$. 
It is interesting to note that the radial motion is governed by 
the same Hamiltonian as in~\cite{deAlfaro:1976je,Pioline:2002qz}, 
and therefore exhibits one-dimensional conformal invariance\footnote{This
is not to be confused with the putative conformally invariant 
boundary quantum mechanics.}

The motion along $r$ is easily integrated in the gauge $N=1$ to 
\be
\label{eqrrho}
r=\frac{C}{\sinh(C\tau)}\ ,\quad \rho=\frac{C}{\tanh(C\tau)}\ .
\ee
By looking at the behavior of the metric near $\tau=\infty$, it is
easy to see that the integration constant $C$ is related to the 
Hawking temperature $T_H$ and black hole entropy $S_{BH}$ 
through~\cite{Ferrara:1997tw}
\be
C=2 S_{BH}T_{H} \ .
\ee
Non-extremal black holes have $C>0$ (the opposite sign results
in a naked singularity), while extremal black holes correspond 
to light-like geodesics\footnote{This is a necessary condition only,
in general one must also fine-tune the velocities at infinity
in order to ensure a smooth solution \cite{Tripathy:2005qp}.}, with $C=0$. 
In this case, the first
and last term in \eqref{hwdw} must cancel, 
\be
\label{rnsusy}
r'=N \ ,
\ee
leading to flat spatial slices $\Sigma$. In the gauge $N=1$,
Equations \eqref{eqrrho} imply that the affine parameter
is the inverse of the radial distance, $\tau=1/r=1/\rho$.
While one may dispose of the radial variable $r$ altogether, it is
however advantageous to retain it for the purpose of defining observables 
such as the horizon area, $A_H = 4\pi e^{-2U} r^2\vert_{U\to -\infty}$
and the ADM mass $m= r (e^{2U}-1)\vert_{U\to 0}$.

As anticipated in~\eqref{heis}, 
the isometries of ${\cal M}_3^*$ lead to conserved 
Noether charges,
\bea
\label{consca}
q_I \ d\tau&=&e^{-2U} \left[ (\Im{\cal N})_{IJ}
d\zeta^J +(\Re {\cal N})_{IJ} (\Im{\cal N})^{ JL}
\left( d\tilde{\zeta}_L + (\Re {\cal N})_{LM} d\zeta^M \right) \right] 
-2 k\, \tilde \zeta_I \, ,\nn
\\
p^I  \ d\tau&=&e^{-2U} ( \Im{\cal N})^{ IL}
\left(d\tilde{\zeta}_L + (\Re {\cal N})_{LM} d\zeta^M \right)
+2 k\, \zeta^I \, ,\label{conscha}
\\
k \ d\tau\ &=& 2\,e^{-4U} \left(d\sigma-\zeta^I 
d\tilde \zeta_I + \tilde \zeta^I    d\zeta_I\right)\, ,\nn
\eea
identified as the electric, magnetic and NUT charges
$p^I,q_I,k$. Their Poisson brackets of course obey the 
same algebra as the Killing vectors \eqref{heis}.

The NUT charge $k$ is related to the 
off-diagonal term in the metric~\eqref{stationan} via 
$\omega=k\cos\theta\ d\phi$. When $k\neq 0$, the metric 
\be
ds_4^2 = -e^{2U} (dt + k\, \cos\theta\, d\phi)^2  + e^{-2U}
[d\rho^2 + r^2 (d\theta^2 + \sin^2\theta\, d\phi^2)]\, ,
\ee
has closed timelike curves along the compact $\phi$ coordinates 
near $\theta=0$, all the way from 
infinity to the horizon. Bona fide black holes have $k=0$, which
corresponds to a ``classical'' limit of the Heisenberg algebra~\eqref{heis}.

Using the conserved charges~\eqref{consca},
one may express the Hamiltonian 
for affinely parameterized geodesic motion on ${\cal M}_3^*$ as
\be
\label{hampqk}
H \equiv p_m g^{mn} p_n = \frac14 p_U^2 + \frac12 p_{z^i} g^{ij} 
p_{z^j} - e^{2U} V_{BH} + \frac14 k^2 e^{4U} \ ,
\ee
where $p_U,p_{z^i}$ are the momenta canonically conjugate to $U,z^i$,
\be
V_{BH}(p,q,z) = -\frac12 (\hat q_I - (\Re\CN)_{IJ} \hat p^J)
(\Im\CN)^{IK} (\hat q_K - (\Re \CN)_{KL} \hat p^L)
-\frac12 \hat p^I (\Im\CN)_{IJ} \hat p^J\, ,
\ee
and
\be
\hat p^I = p^I -2 k \zeta^I\ ,\quad \hat q_I = q_I +2 k \tzeta_I\ .\quad 
\ee
Following~\cite{Ferrara:1997tw}, we refer to $V_{BH}$ 
as the ``black hole potential'', keeping in mind that
it contributes negatively to the actual 
potential governing the Hamiltonian motion $V=-e^{2U}V_{BH}+k^2 e^{4U}$. 
For $k=0$, the motion along 
$(\zeta^I,\tzeta_I,\sigma)$ separates from that along $(U,z^i)$, effectively
producing a potential for these variables.
The attractor flow equations, 
to be discussed in Section \ref{cmap} below, correspond to the restricted
class of supersymmetric solutions to~\eqref{hampqk}.

\subsection{Radial Quantization of Spherically Symmetric 
Black Holes\label{sphgeoq}}
Having shown the equivalence between the radial evolution equations
for stationary, spherically symmetric geometries and the geodesic
motion of a fiducial particle on the cone ${\cal C}$ over ${\cal M}_3^*$, 
quantization is now in principle straightforward:
replace functions on the classical phase space $T^*({\cal C})$
by square integrable wave functions on ${\cal C}$, 
satisfying mass-shell (Wheeler-De Witt) condition
\be
\label{kleing}
\left[-\frac{\pa^2}{\pa r^2} + \frac{\Delta_3}{r^2} -1 \right] 
\Psi(r, U, z^i; \zeta^I, \tzeta_I, \sigma )  = 0\ .
\ee
Here, $\Delta_3$ is the Laplace-Beltrami operator on ${\cal M}_3^*$,
(the quantum analogue of the Hamiltonian $-4H$) 
\bea
\label{lapbel}
\Delta_3&=&
\pa_U^2 + \Delta_4 + e^{4U} \pa_\sigma^2
+2 e^{2U} 
\left[ (\Im{\cal N})_{IJ} \, \nabla^I \, \nabla^J \right.\\
&&\left. + (\Im{\cal N})^{IJ} 
\left( \nabla_I - (\Re{\cal N})_{IK} \nabla^K \right)
\left( \nabla_J - (\Re{\cal N})_{JL} \nabla^L \right) \right]\, ,\nn 
\eea
while $\Delta_4$ is the  Laplace-Beltrami operator on 
the four-dimensional moduli space $\cM_4$, 
\be
\nabla_{I}= \pa_{\zeta^I}- \tzeta^I \pa_\sigma ,\quad 
\nabla^I = \pa_{\tzeta_I} +\zeta^I \pa_\sigma\, ,
\ee
and we have rescaled the wave function $\Psi$ with appropriate powers
of $r$ and $e^{U}$ to cancel the $\pa_r$ and $\pa_U$ linear
derivatives in the above equations.

The wave equation
separates into a Bessel-type equation for the radial direction $r$
and a Laplace equation along ${\cal M}_3^*$:
\be
\label{sepr}
\Psi(r, U, z^i; \zeta^I, \tzeta_I, \sigma ) = \sqrt{r} \left[
\alpha\ J_{\frac12\sqrt{1-4C^2}}(r) + 
\beta\ Y_{\frac12\sqrt{1-4C^2}}(r) \right]\ 
\Psi_{C}( U, z^i, \zeta^I, \tzeta_I, \sigma)\, ,
\ee
where 
\be
\label{Depsi3}
\left[ \Delta_3  + C^2 \right]
\ \Psi_{C}( U, z^i, \zeta^I, \tzeta_I, \sigma) = 0\, .
\ee
In practice, we may also be interested in wave functions which are 
eigenmodes of the electric and magnetic charge operators, given by
the differential operators in \eqref{killingh},
\be
\Psi_{C}(U, z^i, \zeta^I, \tzeta_I, \sigma )  = \Psi_{C,p,q}(U, z^i)\ 
e^{i (p^I \tzeta_I+ q_I \zeta^I)} \, ,
\ee
which is then automatically a zero eigenmode of the NUT charge $k$.
Note however that, due to the Heisenberg algebra \eqref{heis}, 
it is impossible to simultaneously diagonalize the
ADM mass operator $M$, unless either $p^I$ or $q_I$ vanish. Equation
\eqref{Depsi3} then implies that the wave function  
$\Psi_{C,p,q}(U, z^i)$ should satisfy the quantum version of
\eqref{hampqk},
\be
\label{qHampqk}
\left[ \pa_U^2 + \Delta_4 +4 e^{2U} V_{BH}(p,q,z) + C^2 \right]
\ \Psi_{C,p,q}(U, z^i)=0\, .
\ee
The wave function $\Psi_{C,p,q}(U, z^i)$ is
the main object of interest in this paper, and describes the 
quantum fluctuations of the scalars $z^i$ as a function of the scale $e^{U}$
of the time direction ({\it i.e.} effectively as a function of the distance to
the horizon). Alternatively, one may study the full wave function
$\Psi_{p,q}(r,U,z^i)$ as a function of the radius $r$: changing variable
from $r$ to $A_H=4\pi r^2 e^{-2U}$ gives 
access to the quantum fluctuations of the horizon area $A_H$.
In the absence of supersymmetry, it is hardly surprising 
that the wave function is not uniquely specified
by the charges and extremality parameter, as  the condition~\eqref {qHampqk} 
leaves an infinite dimensional Hilbert space.  

An important aspect of quantization is the definition
of an inner product: as in similar instances of mini-superspace 
quantization, the $L^2$ norm on the space of functions on ${\cal C}$
is inadequate for defining expectation values, since it involves
an integration along the ``time'' direction $r$ at which one is supposed
to perform measurements. The customary solution to this problem is
to note that~\eqref{kleing} is a Klein-Gordon-type equation,
and to replace the $L^2$ norm on ${\cal C}$ by the 
$r$-independent Wronskian
\be
\label{kg1}
\langle\Psi | \Psi \rangle =\int dU \ dz^i \ d\zeta^I \ d\tzeta_I \ d\sigma
\, e^{-2(n_V+2)U}\sqrt{\det(g_{ij})}\,  
\Psi^* \stackrel{\leftrightarrow}{\pa_{r}} \Psi \ .
\ee
For factorized wave functions~\eqref{sepr}, the resulting norm is
proportional to the $L^2$ norm on ${\cal M}_3^*$. 
A severe malady of this construction is that
the above scalar product is not positive definite. The standard 
remedy is
to perform a ``second quantization'' and replace the wave function $\Psi$
by an operator; a similar procedure can be followed here, in analogy
with ``third quantization'' in quantum cosmology~\cite{Giddings:1988wv}. 
It is reasonable to
expect that this procedure describes multi-centered geometries.
Fortunately, as we shall see in the next Section, the situation 
is much improved for BPS states, since the Klein-Gordon product~\eqref{kg1} is
(formally) positive definite when restricted to this sector.

\section{BPS Black Holes in $\cN=2$ Supergravity and Twistors\label{sqm}}
We now specialize to supersymmetric black holes in
$\CN=2$ supergravity. In Subsection \ref{cmap}, we review the \qk
geometry of the resulting ${\cal M}_3^*$, identify the geodesics
which correspond to black holes preserving half of the
supersymmetries, and recover the known form of the attractor equations.
In Subsection \ref{twisec}, we outline the construction of the
twistor space $\cZ$ and Swann space $\cS$ over ${\cal M}_3$.
These provide the most convenient framework to formulate and solve the 
BPS conditions. In Subsection \ref{bpsphase}, we show that the phase space of 
BPS black holes is isomorphic to the twistor space $\cZ$, and that BPS
black holes correspond to holomorphic geodesics on $\cS$.
Finally, in Subsection \ref{secpen} we propose a quantization
scheme for spherically symmetric BPS configurations, based on the
Penrose transform between cohomology classes valued in a certain holomorphic line
bundle on $\cZ$ and solutions of certain second order partial differential
equations on the \qk base ${\cal M}_3$. In this framework, we obtain
the exact wave function for a BPS black hole with fixed electric and
magnetic charges, and discuss some of its properties. While most of 
the mathematical results in this Section were obtained in 
\cite{Neitzke:2007ke}, our aim here is to illuminate the physics
motivations behind these mathematical constructions.

\subsection{Attractor Flow and Geodesic Flow\label{cmap}}
Four-dimensional 
$\CN=2$ supergravity with $n_V$ vector multiplets consists of
$n_S/2=n_V$ complex scalars, $n_A=n_V+1$ Maxwell fields (including
the graviphoton), two gravitini and $n_V$ gaugini (hypermultiplets 
may be safely ignored as they are not sourced by black holes).  
The couplings in the four-dimensional action \eqref{4dact} 
are determined in terms of a holomorphic
prepotential function $F(X^I)$. The manifold ${\cal M}_4$
is a projective special K\"ahler manifold with K\"ahler potential
\be
\label{kahl}
{\cal K}(z^i,\bar z^{\overline j}) = -\log K(X,\bar X) 
= - \log \left[ i \left( \bar X^I F_I - X^I \bar F_I\right) \right]\ ,
\ee
where $F_I:=\pa F(X)/\pa X^I$, $I=0\dots n_V$,
while the gauge kinetic terms are related to 
the second derivative $\tau_{IJ}:=\pa_{I}\pa_{J}F(X)$ via
\be
{\cal N}_{IJ} = \bar\tau_{IJ} + 2i \frac{ (\Im\tau\cdot X)_I ~
(\Im\tau\cdot X)_J}{X\cdot\Im\tau\cdot X} \ .
\ee

The scalar manifold ${\cal M}_3$ obtained by Kaluza-Klein reduction to
three dimensions is a quaternionic-K\"ahler space, obtained by the 
``c-map'' from the special K\"ahler manifold ${\cal
M}_4$ \cite{Gunaydin:1983rk,Gunaydin:1983bi,Cecotti:1988qn,Ferrara:1989ik}.  The analytically
continued ${\cal M}_3^*$ is a para-quaternionic-K\"ahler space, which we
shall refer to as the c$^*$-map of ${\cal M}_4$.  While
${\cal M}_3$ has a Riemannian metric with special holonomy
$USp(2)\times USp(2n_V+2)$, ${\cal M}_3^*$ has a split signature
metric with special holonomy $Sp(2,\R)\times Sp(2n_V+2,\R)$.  As mentioned in 
the introduction, we work for convenience with the more familiar Riemannian 
space ${\cal M}_3$, leaving the analytic continuation implicit most of the time.

In addition to the bosonic fields appearing in \eqref{clasme}, the 
three-dimensional Lagrangian contains also the fermionic partners of $\phi^m$
and of the graviton, resulting in $\cN=4$ Euclidean supergravity
in three dimensions. Upon further restriction to spherically symmetric
solutions, one expects to find fermionic partners for the one-dimensional 
graviton $N$ and the bosonic fields $r,\phi^m$ in 
$\IR^+ \times {\cal M}_3^*$, such that the resulting Lagrangian has
$\cN=4$ supersymmetry in one dimension\footnote{Note that a 
spherically covariant Killing spinor
in three dimensions decomposes as $\epsilon^{A'}_\alpha = 
\eps^{A'}(\rho)\chi_\alpha$ where $\chi_\alpha$ is a Killing 
spinor on $S^2$. As a result, the number of supercharges is halved
by the spherical reduction.}. The resulting one-dimensional
supergravity model will be presented in \cite{w-toappear}. 
For the present purposes,
we only require the supersymmetry transformations of the fermions, the
reduction of which is given in Appendix A. 
To describe this explicitly, let us 
recall some basic features of quaternionic-K\"ahler manifolds.
The restricted holonomy
implies that the complexified tangent bundle of  ${\cal M}_3$ splits locally
as 
\be
\label{teh}
T_\IC {\cal M}_3 = E \otimes H\ ,
\ee
where $E$ and $H$ are complex vector bundles of respective dimensions 
$2n_V+2$ and $2$. This decomposition is preserved by the 
Levi-Civita connection. The latter decomposes into
its $USp(2)$ and $USp(2n_V+2)$ parts $p$ and $q$,
\be
\label{omab}
\Omega_{AA'}^{BB'} =  p_{A'}^{B'} \delta_{A}^{B} + q^A_B \delta_{A'}^{B'}\ ,
\ee
where $\epsilon_{A'B'}$, $\eps_{AB}$ are the
antisymmetric tensors invariant under $USp(2)$, $USp(2n_V+2)$ respectively.
The change of basis from $T_\IC {\cal M}_3$ to $E\otimes H$
is achieved by a covariantly constant
``quaternionic vielbein'' $V^{AA'}=V^{AA'}_m d\phi^m$ 
($A=1,..,2n_V+2,\, A'=1,2,\, m=1,\dots, 4n_V+4$), from  which 
one can construct the metric $ds^2$, as well as three almost 
complex structures and their two-forms $\omega^i$,
\be
ds^2 =\epsilon_{A'B'}\ \eps_{AB}\ 
 V^{AA'}   \otimes V^{BB'} \ ,\quad 
\omega^i = \epsilon_{A'B'}\ (\sigma^i)^{B'}_{C'} 
\ \eps_{AB}\ 
 V^{AA'}   \wedge V^{BC'}\ .
\label{metric}
\ee
The fermions in the non-linear $\cN=4$ sigma model on $\cM_3$
transform under $USp(2n_V+2)$ and are $USp(2)$-inert\footnote{In fact,
the $\cN=4$ one-dimensional sigma model is a reduction of the 
original $\N=2$ locally supersymmetric sigma model in four
dimensions~\cite{Bagger:1983tt}.}, 
with supersymmetric variations~\cite{w-toappear} 
\be
\label{dpsi}
\delta \chi^A = -\frac{1}{N} V_m^{AA'} \,  \ophi{}\!^m\  \epsilon_{A'} \ ,
\ee
where $\ophi{}^m$ is the supercovariant time derivative of $\phi^m$,
which reduces to the usual time derivative $\phi^{'m}$ for zero value of the 
worldline gravitino.

From~\eqref{dpsi}, it is apparent that supersymmetric solutions
are obtained when $V^{AA'}$ has a null eigenvector,
\bse
\label{susycond}
\bea
\label{susycond1}
\mbox{SUSY}\quad &\Leftrightarrow & \quad
\exists\,\epsilon_{A'}\ \vert \ 
V^{AA'} \eps_{A'} = 0 \\ 
\label{susycond2}
&\Leftrightarrow &\quad 
\epsilon_{A'B'}\,V^{AA'} V^{BB'} = 0\ .
\eea
\ese
For fixed $\epsilon^{A'}$, these are $2n_V+2$ conditions on 
the velocity vector $\phi^{'m}$ at any point along the geodesic,
removing half of the degrees of freedom from the generic trajectories.
We now demonstrate that these conditions imply the
usual attractor flow equations generalized to include the 
NUT charge.

For the case of the $c$-map ${\cal M}_3$, the quaternionic vielbein 
was computed explicitly in~\cite{Ferrara:1989ik}. After analytic
continuation, one obtains
\be
V^{AA'} = \begin{pmatrix} i u & v \\ e^a & i E^a  \\
-i \bar E^{\bar a} & \bar e^{\bar a} \\ -\bar v & i \bar u\end{pmatrix} \ .
\ee
where $e^a=e^a_i dz^i$ is a vielbein of the special K\"ahler manifold, 
$e^a_i \bar e_{\bar a\overline \jmath} \delta_{a\bar a}
=g_{i\overline \jmath}$, and
\bse
\bea
u&=& e^{K/2-U} X^I \left( d\tilde\zeta_I 
+{\cal N}_{IJ}d\zeta^J \right) \ ,\\
v&=&  dU- i\, e^{-2U}\left( 
  d\sigma-\tilde\zeta^I d \zeta_I+\zeta^Id\tilde\zeta_I\right) \ ,\\
E^a&=& e^{-U} e^{a}_{i} g^{i\overline \jmath} \bar f_{ \overline \jmath}^I 
\left( d\tilde\zeta_I + {\cal N}_{IJ}d\zeta^J \right)\label{vb}\ .
\eea
\ese
Expressing $d\zeta^I, d\tilde\zeta_I, d\sigma$ in terms of the conserved
charges~\eqref{conscha}, the entries in the quaternionic vielbein may 
be rewritten as
\bse
\bea
u&=& i e^{K/2 + U} X^I \left[ q_I -  2 k \tilde\zeta_I 
-{\cal N}_{IJ} (p^J + 2 k \zeta^J)  \right] d\tau \ ,\quad\\
v&=&  dU- {i\over 2} e^{2U} k \ d\tau\ ,\\
e^a &=&  e^a_i\ d z^i\ ,\quad\\
E^a&=& i\,e^{U} e^{a i} g^{i\overline \jmath} \bar f_{ \overline \jmath}^I 
\left[ q_I - 2k \tilde\zeta_I -  {\cal N}_{IJ} (p^J + 2k \zeta^J) \right] 
d\tau\, .
\eea
\ese
Now we return to the supersymmetry variation of the fermions~\eqref{dpsi}:
the existence of $\eps^{A'}$ such that 
$\delta\chi^{A}$ vanishes implies that the first column of $V$ has to
be proportional to the second, hence
\bse
\bea
\frac{dU}{d\tau}- {i\over 2} e^{2U} k &=& 
-i\ e^{i\theta}\ e^{K/2+U}\ X^I \left( q_I - 2k \tilde\zeta_I 
-{\cal N}_{IJ} (p^J + 2k \zeta^J)  \right) \, ,\\
\frac{d z^i}{d\tau} &=& -i\ e^{i\theta} \ e^{U}\ g^{i\overline \jmath} \bar f_{ \overline \jmath}^I 
\left( q_I - 2k \tilde\zeta_I -  {\cal N}_{IJ} (p^J + 2k \zeta^J) \right)\, ,
\eea
\ese
where the phase $\theta$ is determined by requiring the reality of $U$. 
For vanishing NUT charge, this becomes
the well-known attractor flow equations ~\cite{Ferrara:1995ih,
Ferrara:1996um,Ferrara:1997tw,Moore:1998pn,Denef:2000nb} 
\bse
\label{att}
\bea
\frac{dU}{d\tau} &=& - e^{U}\ |Z| \ , \label{att1}\\
\frac{dz^i}{d\tau} &=& - 2 ~e^U~ g^{i\overline \jmath} 
\pa_{\overline \jmath} |Z|\ , \label{att2}
\eea
\ese 
where $Z$ is the central charge
\be \label{centca}
Z(p,q;z^i,\bar z^{\bar i})
= e^{\cK/2} \left( p^I F_I - q_I X^I \right) \ .
\ee
The equivalence between 
the attractor flow equations on ${\cal M}_4$
and supersymmetric geodesic motion on ${\cal M}_3$ 
was observed long ago in~\cite{Gutperle:2000ve}, and is
a consequence of the T-duality between black holes and 
instantons~\cite{Behrndt:1997ch,deVroome:2006xu,npv-to-appear}.

Having reproduced the usual form of the attractor equations,
we return to the supersymmetry conditions \eqref{susycond},
and comment on their structure. The quaternionic viel-bein
$V^{AA'}/d\tau=V^{AA'}  d\phi^m/d\tau$ can be viewed as 
a $2 \times (2n_V+2)$ matrix of functions on the unconstrained phase space
$T^*({\cal M}_3)$, after expressing the velocity  $d\phi^m/d\tau$
in terms of the momentum $p_m$. Similarly, the quadratic constraints
\be
\label{quadbps}
H_{AB}\equiv  \epsilon_{A'B'}\,V^{BB'}\, V^{AA'} / d\tau^2  = 0\, ,
\ee
are functions on the unconstrained phase space, corresponding
to the $2\times 2$ minor determinants of the matrix $V^{AA'}$. The
constraints $H_{AB}\equiv 0$ are first class, in the sense that
their Poisson brackets vanish on the constrained locus. Indeed,
computations show that
\be
[H_{AB},H_{CD}]=-8\,V_{[AA'}\, q^{A'}_{B]}{}^E_{[C}\, H_{D]E}\, .
\ee
where $q^{A'E}_{BC}$ is the $USp(2n_V+2)$ connection, whose one-form index
has been traded to $A'E$ using the inverse of the quaternionic vielbein.
The constraints $H_{AB}$ are not independent however, since the rank one
condition on $V^{AA'}$ enforces only $2n_V+1$ conditions on its 
$4n_V+4$ entries. Since each first class constraint reduces the dimension
by two, the real dimension of the BPS phase space is $8n_V+8-2(2n_V+1)=4n_V+6$.
The symplectic structure on this space is however obscure from this
construction. In the next Section, we show that  once the Killing 
spinor $\epsilon^{A'}$ is included, the BPS phase space is realized as the
twistor space $\cZ$ of $\cM_3$, with complex dimension $2n_V+3$.

\subsection{Twistor Space and Swann Bundle\label{twisec}}
The one-dimensional ${\cal N}=4$ non-linear sigma model on $\cM_3$
is unusual because the three complex structures responsible for extended 
supersymmetry are not integrable.
This is hardly surprising because the model must also be coupled to worldline 
gravity. Exactly such a study is underway~\cite{w-toappear},
however, for BPS configurations, this problem can also be circumvented
by a standard mathematical construction which physically incorporates
the Killing spinor in the black hole geometry, as we discuss further in
Subsection \ref{bpsphase}. 

Let $\cS$ be the total 
space\footnote{More precisely, $\cS$ is the total space of 
$H^\times /\IZ_2$, where $H^\times$ is the bundle $H$ with the zero
section deleted and $\IZ_2$ acts as $\pi^{A'}\to -\pi^{A'}$ on the fiber of 
$H$.} of the bundle $H$ over ${\cal M}_3$. This $4n_V+8$ 
dimensional space, known as the Swann bundle or \hk cone, admits a 
dilation and $SU(2)$-invariant \hk metric~\cite{MR1096180,deWit:2001dj}
\be
\label{dssw}
ds^2_{\cS} =|D\pi|^2 + \frac{\nu}{4} \,R^2 \,ds^2_{{\cal M}_3}\, .
\ee
Here, $\pi^{A'}$ are coordinates in the $\IR^4$ fiber of $H$, 
$R^2=|\pi^1|^2 + |\pi^2|^2$ is the $USp(2)$ invariant norm,
and $D\pi^{A'}$ is the covariant exterior derivative of $\pi^{A'}$,
\be
\label{DPi}
D\pi^{A'} = d\pi^{A'} + p^{A'}_{B'} \pi^{B'}\ ,
\ee
and $\nu$ is related to the scalar curvature of the base by
$R=4n(n+2)\nu$. In particular, $\nu<0$, and $\cS$ has quaternionic
Lorentzian signature $(1,n_V+1)$ and holonomy 
$USp(2,2n_V+2)$. The spin connection $\Omega^{\aleph}_{\beth}$ and the
covariantly constant quaternionic vielbein ${\cal V}^{\aleph}$ 
(where $\aleph \in\{A,A'\}$ runs over two more indices than $A$)
can be simply obtained from the quaternionic vielbein $V^{AA'}$ on 
the base $\cM_3$ via
\be
\label{vsw}
{\cal V}^{\aleph} = 
\begin{pmatrix} 
D\pi^{A'} \\ \hline  V^{AA'} \pi_{A'}
\end{pmatrix} \ ,\quad
\Omega^{\aleph}_{\beth} =\left(
\begin{array}{c|c}
p^{A'}_{B'} & V^{AA'} \\ \hline V_{BB'} & q^A_{B}
\end{array}\right)\ .
\ee
The vielbein ${\cal V}^{\aleph}$ gives a set of $(1,0)$-forms on $\cS$
(for a particular complex structure), which together with 
$\bar{\cal V}$ span the cotangent space of $\cS$. 

\begin{figure}
\centerline{\hfill
\includegraphics[height=7cm]{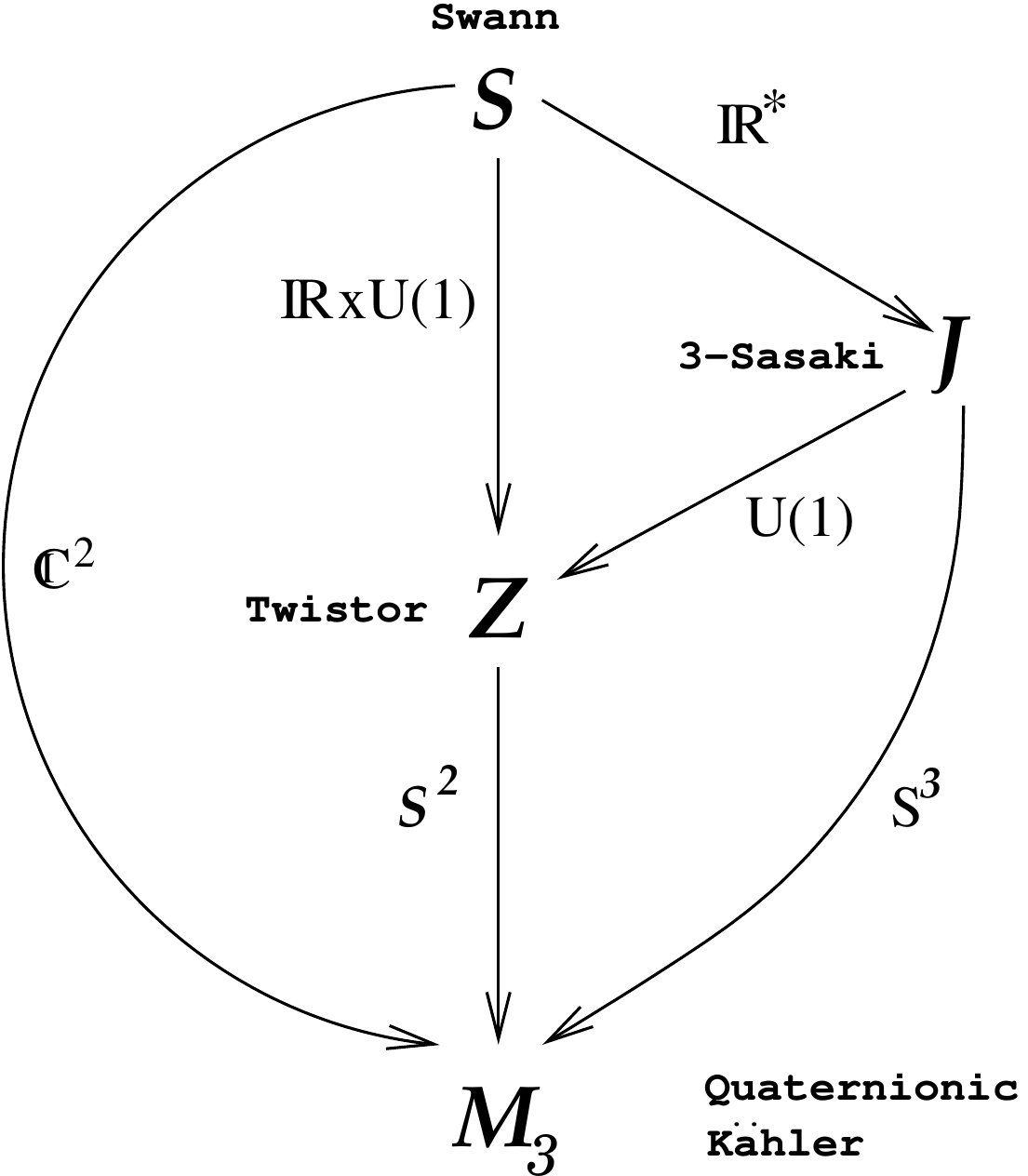}
\hfill}
\caption{Bundles over the \qk space $\cM_3$.\label{hkcfig}}
\end{figure}

It is useful to view the unit sphere $S^3$ in $H$ as a Hopf fibration
and choose coordinates
\be
e^{i\varphi} =\sqrt{\pi^2/\bar \pi^2}\ ,\quad
z=\pi^1/\pi^2\, ,
\ee
on the $U(1)$ fiber and $S^2$, respectively. 
The \hk cone metric \eqref{dssw} can then be rewritten as
\be
\label{dssw2}
ds^2_{\cS} = d R^2 + R^2 \left( 
\sigma_1^2+\sigma_2^2+\sigma_3^2+
\frac{\nu}{4} ds^2_{\cM_3} \right) \ ,
\ee
where the triplet of 1-forms
\be
\sigma_1+i \sigma_2 = \frac{dz+{\cal P}}{1+z\bar z}\ ,\quad
\sigma_3 =  d\varphi - \frac{i}{2(1+z\bar z)}
(\bar z dz - z d\bar z) - \frac{i}{r^2} \pi^{A'} p_{A'}^{B'}\bar \pi_{B'}\, ,
\ee
and $\cal P$ is the projectivized USp(2) connection,
\be
{\cal P}=p_2^1 + z(p^1_1-p^2_2) - z^2 p_1^2\ .
\ee
Hence $\cS$ is a real cone  
over a  $4n_V+7$-dimensional 3-Sasaki space $\cJ$, which in turn
is a $U(1)$ bundle over a  
$4n_V+6$-dimensional ``twistor'' space $\cZ$ with metric 
\be
\label{dsz}
ds^2_{\cZ}=\sigma_1^2+\sigma_2^ 2+ \frac{\nu}{4} 
ds^2_{\cM_3}=\frac{|dz + {\cal P}|^2}{(1+\bar z z)^2}
+ \frac{\nu}{4} ds^2_{\cM_3}\, .
\ee
The twistor space $\cZ$ is an $S^2$ bundle over ${\cal M}_3$, 
with complex Lorentzian signature $(1,2n_V+2)$, see Figure \ref{hkcfig}.
The twistor space can also be obtained
from $\cS$  directly as the K\"ahler quotient
by the $U(1)$ symmetry shifting the coordinate $\varphi$ 
(at unit value of the moment map $|\pi|^2$).
In particular, it carries a canonical complex structure 
whose K\"ahler form is 
\be
\omega_\cZ = i\frac{|dz + {\cal P}|^2}{(1+\bar z z)^2} -
\frac{i \nu}{2(1+z\bar z)} 
\left[ (z+\bar z) \omega^1 
+ i (z-\bar z) \omega^2 + (1-z\bar z) \omega^3 \right]\ ,
\ee
where $\omega_i$ are the quaternionic 2-forms in \eqref{metric}.

Isometries on $\cM_3$ lift to holomorphic isometries on 
$\cZ$ \cite{MR872143,deWit:2001bk}, and tri-holomorphic isometries on $\cS$. 
A set of complex coordinates $\xi^I,\txi_I,\alpha$ on the Swann 
bundle $\cS$ and 
twistor space $\cZ$ adapted to the Heisenberg symmetries was constructed
in \cite{Neitzke:2007ke}. In terms of these coordinates, 
the complexified Heisenberg algebra acts as
\be
\label{cxkillingh}
P^I= \pa_{\txi_I} - \xi^I \pa_\alpha\ ,\quad
Q_I = -\pa_{\xi^I} - \txi_I \pa_\alpha\ ,\quad
K = \pa_\alpha\, .
\ee
Only the real Heisenberg algebra $P^I+\bar P^I, Q_I+\bar Q_I, K+\bar K$
is an isometry of $\cZ$, however. The K\"ahler-Einstein metric on $\cZ$ 
may be obtained from the K\"ahler potential \cite{Neitzke:2007ke}
\be
\label{kz-hesse}
K_{{\cal Z}}= \frac12\log\left\{
 \Sigma^2 \left[\frac{i}{2}(\xi^I-\bar\xi^I),
\frac{i}{2}(\txi_I-\bar\txi_I)\right]
+\frac{1}{16} \left[
\alpha -\bar\alpha+ \xi^I \bar\txi_I-\bar\xi^I \txi_I
\right]^2 \right\}+\log 2\ ,
\ee
where $\Sigma(\phi^I,\chi_I)$ is the Hesse potential associated to the
special geometry of the four-dimensional moduli space; namely, the
Legendre transform of the ``topological free energy'' with respect to
the magnetic charge $p^I$ \cite{Hitchin:2005uu,LopesCardoso:2006bg,
Ferrara:2006js},
\be
\Sigma(\phi^I,\chi_I) = \langle\ 
\frac{1}{2i}\left[ F(p^I + i \chi^I) - \bar F(p^I - i \chi^I) \right]
+ i p^I \chi_I \rangle_{p^I}\, .
\ee
Note in particular that $\Sigma$ has the same 
functional dependence on the ``potentials'' $(\phi^I,\chi_I)$ 
as the tree-level black hole entropy on the charges $(p^I,q_I)$,
and is invariant under symplectic rotations of  $(\phi^I,\chi_I)$.

The complex coordinates on $\cS$ can be obtained by supplementing 
$(\xi^I,\txi_I,\alpha)$ with one complex coordinate $v^\flat$
such that $U(1)$ acts by rotating the phase of $v^\flat$, and 
\be
R^2 = |v^\flat|^2 \ e^{K_{\cZ}}\, .
\ee
This quantity $\chi=R^2$ in fact equals the hyperk\"ahler potential 
of $\cS$, a simultaneous 
K\"ahler potential for its two-sphere's worth of complex structures.

The relation between the complex coordinates $\xi^I,\txi_I,\alpha$
(and their complex conjugates) on $\cZ$ 
and the coordinates $U,z^i,\zeta^I,\tzeta_I,
\sigma$ on the \qk base as well as the fiber coordinate $z\in\C\PP^1$
was worked out in  \cite{Neitzke:2007ke} by 
forming $\IR^{\times} \times SU(2)$ invariants, leading to the ``twistor map''
\bse
\label{gentwi}
\bea
\xi^I &=& \zeta^I + 2i\ e^{U+{\cal K}(X,\bar X)/2} 
\left( z \bar X^{I} + z^{-1} X^{I} \right)\ , \label{gentwi1}\\
\txi_I &=& \tzeta_I + 2i\ e^{U+{\cal K}(X,\bar X)/2}  
\left( z\ \bar F_I +  z^{-1}\ F_I \right)\ , \label{gentwi2}\\
\alpha &=& \sigma + \zeta^I \txi_I -\tzeta_I  \xi^I\ ,
\label{gentwi3}
\eea
\ese
where $(X^I,F_I)$ and ${\cal K}(X)$ have been defined in Section \ref{cmap}.
In the para-quaternionic case relevant for black holes, the holomorphic
and anti-holomorphic variables become independent real
variables. This may be further lifted to $\cS$ by using
\begin{equation} \label{hkc-coords}
\begin{pmatrix} \pi^1 \\ \pi^2 \end{pmatrix} = 2\,e^U\, \sqrt{v^\flat} 
\begin{pmatrix} z^\half \\ z^{-\half} \end{pmatrix}.
\end{equation}
A key feature is that, for a fixed point on the base, the complex
coordinates $\xi^I,\txi_I,\alpha$ depend rationally on the coordinate
$z$ in the twistor fiber; said differently, the fiber over
any point on the base is rationally embedded in $\cZ$.
This distinctive property of twistor spaces is the origin
of the Penrose transform between holomorphic sections of $\cO(-\ell)$
on $\cZ$ and harmonic sections on $\cM_3$. We shall return to this
topic in Section~\ref{secpen}.

\subsection{The BPS Phase Space as the Twistor Space\label{bpsphase}}

We now return to physics and show that supersymmetric black holes
correspond to a special class of geodesics on  $\cM_3$ which
can be lifted holomorphically to the Swann space $\cS$.
We emphasize again (see Footnote \ref{para-fn} of the introduction) that 
$\cM_3$ and related spaces are complexified, despite our 
use for convenience of the language appropriate to the real slice $\cM_3$.

First we observe that geodesic motion on $\cM_3$ is equivalent to
geodesic motion on $\cS$, provided one restricts to trajectories with
vanishing angular momentum along $S^3$. Indeed, geodesic motion on $\cS$
decouples into a radial motion along $R$, with a conformal Hamiltonian
of the same type as in \eqref{rrhoc}, and geodesic motion along 
the 3-Sasakian base $\cJ$. The restriction to zero angular momentum along
$S^3$ can be enforced by gauging the $SU(2)$ isometries $\sigma_i \to 
\sigma_i + {\cal A}_i$, and restricting to the $SU(2)$-singlet sector.

Similarly, the one-dimensional $\cN=4$ non-linear sigma model on $\cM_3$ should  
be obtained by gauging a $\cN=4$ non-linear sigma model on 
the Swann space $\cS$, with fermions $\psi^{\aleph}$ 
now transforming under $USp(2,2n_V+2)$.
As $\cS$ is \hk, its $USp(2)$ curvature vanishes, so that
the supersymmetry variations on $\cS$ split into holomorphic
and antiholomorphic parts,
\be
\label{dpsisw}
\delta\psi^{\aleph} = \cV^\aleph \epsilon\ ,\quad 
\delta\bar\psi^{\bar \aleph} = 
\bar\cV^\aleph \bar \epsilon \ .
\ee
where $\cV^\aleph$ is the holomorphic vielbein introduced in \eqref{vsw}.
Taking advantage of the $SU(2)$ symmetry on $\cS$, which rotates
$\eps$ into $\bar \eps$, we can assume that the unbroken symmetry
generator is $\bar\eps$. Thus, we could define 
supersymmetric geodesics on $\cS$ as
those whose momentum is purely holomorphic at any point along the
trajectory, namely $\bar{\cal V}^{\bar \aleph}=0$. Using \eqref{vsw}, this
condition may be rewritten as
\be
\label{bpscondhk}
{\rm BPS} \quad \Leftrightarrow \quad 
\left\{ \begin{matrix} D\pi^{A'} &=& 0\, , \\ V^{AA'} \pi_{A'}&=&0\, .
\end{matrix}\right. 
\ee

Now let us compare these conditions with the conditions defining BPS
black holes.
Upon identifying the coordinate in the fiber $\pi^{A'}$ with the 
supersymmetry parameter~$\eps^{A'}$, we recognize the first equation
in \eqref{bpscondhk} as the condition \eqref{susycond} 
for supersymmetric motion on the quaternionic-K\"ahler base. 
In Appendix \ref{redferm}, we show that
the radial dependence of the Killing spinor preserved by the black hole 
solution is indeed governed by $D\eps^{A'}=0$, consistently with the
second equation in \eqref{bpscondhk}.  
Thus, we may identify the $\IR^4$ fiber of the Swann bundle as
the Killing spinor preserved by the black hole geometry.\footnote{Another
way to see that $R^2=|\pi|^2$ is unrelated to the cone coordinate $r$ 
on $\cC$ is that
supersymmetric geodesic motion on $\cS$ is necessarily light-like whereas,
as argued below \eqref{hwdw}, the geodesic motion on $\cC$ 
has to be massive.}
Similarly, the coordinate $z\in\mathbb{P}^1$ on the twistor space $\cZ$
keeps track of the projectivized Killing spinor, $z=\eps^1/\eps^2$.

We conclude from this discussion that stationary, spherically symmetric 
BPS black holes correspond to holomorphic geodesics on the Swann space $\cS$,
with vanishing momentum along the $S^3$ fiber. 
This description will be very useful for the purpose of quantizing
BPS black holes, as explained in the next Subsection. 

This reformulation is already advantageous at the classical level:
in particular, it allows to integrate the BPS equations
of motion explicitly, and recover the known spherically symmetric 
BPS solutions in $\cN=2$ supergravity~\cite{Neitzke:2007ke}.
The key observation is that, as a result of the vanishing of the
anti-holomorphic momenta\footnote{We deviate from \cite{Neitzke:2007ke}
by an overall complex conjugation.} $\bar p_{\bar\aleph}\equiv 
\bar {\cal V}^{\bar\aleph}\equiv 0$,
the holomorphic coordinates $z^{\aleph}$ on $\cS$ are constants
of motion. Moreover, the holomorphic momenta $p_{\aleph}$ are also
constants of motion, related to the conserved charges $p^I,q_I,k$ associated 
to the Heisenberg symmetries. Using the K\"ahler 
property of the metric, the equation $p_{\aleph}=g_{\aleph\bar\aleph} 
dz^{\bar\aleph}/d\tau$ can be integrated into
\be
\label{linflow}
\pa_{z^{\aleph}}\chi = p_{\aleph} \tau + c_{\aleph}\, ,
\ee
where $c_{\aleph}$ are integration constants. This equation may
be solved to express ${\bar z}^{\bar \aleph}$ in terms of the constants
of motion $z^{\aleph}, p_{\aleph}$ and $c_{\aleph}$ and the time $\tau$.
By inverting the twistor map \eqref{gentwi}, the geodesic 
motion on $\cS$ can be projected to the base $\cM_3$.
If we also require that the momentum for the $U(1) \subset SU(2)$ preserving 
the complex structure on $\cS$ should vanish, we recover the
known spherically symmetric solutions of $\cN=2$ supergravity.
We note that (after an appropriate redefinition of $p_{\aleph}$ \cite{Neitzke:2007ke})
the BPS geodesics on $\cM_3$ depend on the constants $z^\aleph$
only via overall shifts of $\zeta^I,\tzeta_I,\sigma$,
corresponding to gauge symmetries in four dimensions. The 
number of physical parameters labeling the solution is therefore $2n_V+4$, or
$2n_V+3$ after enforcing $U(1)$ invariance.

This reformulation also allows us to clarify the geometric nature of the
BPS phase space. In particular, the BPS constraints 
$\bar p_{\bar\aleph}\equiv 0$ are manifestly first class. The BPS
phase space is the symplectic quotient of
the unconstrained phase space $T^*(\cS)$, with symplectic form
$\omega = dz^\aleph\wedge dp_{\aleph} + d{\bar z}^{\bar \aleph}\wedge 
d{\bar p}_{\bar \aleph}$ by the Hamiltonian vector fields
$\pa_{\bar z^{\bar \aleph}}$ associated to these constraints,
corresponding to the afore-mentioned gauge symmetries in four dimensions.
A standard trick to treat first class constraints is to augment them
with gauge fixing constraints, such that the total system is second class.
A simple choice of gauge fixing constraints is to fix the value of
${\bar z}^{\bar \aleph}$ to arbitrary constants, leading to
\be
\omega_{\rm BPS}=  dz^\aleph \wedge dp_{\aleph}\ .
\ee
In this gauge the BPS phase space is the holomorphic
cotangent bundle to $\cS$.  (It would become a real symplectic manifold
if we chose the real slice over $\cM_3^*$.)
Alternatively, the gauge fixing constraint
\be
p_{\aleph} = \gamma \,\pa_{z^\aleph}\chi\ ,
\ee
where $\kappa$ is an arbitrary constant, leads to 
\be
\omega_{\rm BPS}=  \gamma\, \pa_{z^\aleph}\pa_{\bar z^{\bar\aleph}}
\chi\, dz^\aleph \wedge d\bar z^{\bar \aleph} \ ,
\ee
proportional to the K\"ahler form $\omega_{\cS}$ on $\cS$. 
The $U(1)$ invariance can be enforced by performing a further
symplectic quotient. Thus, we may identify the BPS phase space
as the  twistor space $\cZ$, equipped with its K\"ahler form $\omega_\cZ$.
The difference between these two descriptions of the BPS phase space presumably 
arises from singularities in the gauge-fixing conditions, which we have not 
closely investigated.
We note that the value of $\gamma$, irrelevant for local, classical
considerations, becomes important
quantum mechanically, as it determines the normalization of $\omega_{\rm BPS}$ 
and hence the line bundle in which the wave function should be valued.

\subsection{Quantizing Spherically Symmetric BPS Black Holes \label{secpen}}

According to the discussion in Section \ref{cmap} and \ref{bpsphase},
we have two equivalent characterizations of supersymmetric black holes
at our disposal: 
\begin{itemize}
\item[i)] Geodesic motion on the quaternionic-K\"ahler
space ${\cal M}_3$, satisfying the quadratic constraints \eqref{quadbps},
\item[ii)] Holomorphic geodesic motion on the Swann space $\cS$,
with vanishing angular momentum along the $S^3$ in the $\IR^4$ fiber.
\end{itemize}
In the first formulation, it is natural to try and construct
the BPS Hilbert space as a subspace of $L^2(\cM_3)$ annihilated by 
a quantum version of the constraints \eqref{quadbps},
\be
\label{wavebpsqk}
\left[ \eps^{A'B'} \nabla_{AA'}\, \nabla_{BB'} + \kappa\, \eps_{AB} 
\right] \Psi = 0\ .
\ee 
Here, $\nabla_{AA'}=V_{AA'}^{m} \nabla_m$ is the covariant derivative on 
$\cM_3$, and we have allowed for a possible quantum ordering ambiguity 
parameterized by the c-number $\kappa$. While this description 
has the advantage of not introducing any gauge degrees of freedom, 
finding the general solution of the second order partial differential 
system \eqref{wavebpsqk} is a priori difficult.

In the second formulation, the Hilbert space is a priori much simpler
to construct, since the linear supersymmetry conditions \eqref{bpscondhk}
can be quantized as
\be
\label{wavebpssw}
\bar\partial_{\aleph} \Psi = 0\ ,
\ee
where $\bar\partial_{\aleph}$ are partial derivatives with respect
to the antiholomorphic coordinates $\bar z^{\bar\aleph}$ on~$\cS$.
Moreover, the vanishing of the $U(1)$ momentum in the fiber implies
that $\Psi$ should be a homogeneous holomorphic function on $\cS$
of vanishing degree classically. Equivalently, $\Psi$ is 
a holomorphic function on $\cZ$. 

This leads to an immediate 
puzzle:  globally, the only holomorphic 
functions on $\cZ$ are constants.
More care is needed however: in particular, 
we did not include the fermionic degrees of freedom, but
imposed by hand the BPS constraints on the bosonic trajectory.  
Including the fermions and the (super)ghosts in the one-dimensional 
sigma model \eqref{clasme} may lead to a non-zero degree of 
homogeneity $\ell$ on $\cS$, so that $\Psi$ is now a section 
of the line bundle $\cO(-\ell)$ over $\cZ$, 
and possibly replace holomorphic functions by sheaf
cohomology classes, as usual in K\"ahler quantization
 (see {\it e.g.}~\cite{MR94a:58082}).
Since the K\"ahler-Einstein metric
on the twistor space $\cZ$ has two negative eigenvalues,
it is natural to propose\footnote{For very special $\cN=2$ supergravities,
the space $H^1(\cZ,\cO(-\ell))$ for large enough $\ell$
indeed furnishes a unitary representation of $G_3$, belonging
to the quaternionic discrete series~\cite{MR1421947}.}
that $\Psi$ is valued in $H^1(\cZ,\cO(-\ell))$. 
A more detailed analysis of the BRST quantization of the one-dimensional
locally supersymmetric non-linear sigma model on $\cM_3$ is 
left to a forthcoming publication~\cite{w-toappear}. 

Remarkably, there exists a mathematical construction 
valid for any quaternionic-K\"ahler manifold, sometimes known
as the quaternionic Penrose transform \cite{quatman,MR1165872,Neitzke:2007ke}, 
which takes an element of the sheaf cohomology group
$H^1(\cZ,\cO(-2))$ to a solution of the partial differential 
system~\eqref{wavebpsqk}. More generally, the Penrose transform maps
classes in $H^1(\cZ,\cO(-\ell))$ to 
sections $\Psi^{(A'_1 A'_2\dots A'_{\ell-2})}$of $S^{\ell-2}H$, where
$S^{\ell-2}(H)$ is the $(\ell-2)$-fold symmetric power of the rank 2 
bundle $H$ on $\cM_3$ introduced in \eqref{teh}. 

Using the complex coordinate system introduced in Section \ref{twisec}, 
it is easy to provide an explicit integral representation of this transform,
where the element of $H^1(\cZ,\cO(-\ell))$ is represented by a holomorphic
function $g(\xi^I,\txi_I,\alpha)$ in the trivialization 
$v^\flat=1$ \cite{Neitzke:2007ke}:
\begin{equation} \label{penrosetrans}
\Psi^{(A'_1 A'_2\dots A'_{\ell-2})}
(U, z^{i}, \bar z^{\bar j}, \zeta^I, \tzeta_I, \sigma) =
2^\ell\,e^{\ell U}\  \oint \frac{dz}{z}\,z^{\delta/2}
\,g(\xi^I(z), \txi_I(z), \alpha(z))\ .
\end{equation}
where the integer $\delta$ counts the 
number of $i$ such that $A_i'=1$, minus the number of $i$ such that $A_i'=2$,
{\it i.e.} the helicity under $U(1)\subset SU(2)_H$.
In this formula, $\xi^I,\txi_I,\alpha$ are to be expressed as functions
of the coordinates on $\cM_3$ and $z$ via the twistor map \eqref{gentwi}. 
The integral runs over a contour around $z=0$. In 
\cite{Neitzke:2007ke}, it was shown that the left-hand side of
\eqref{penrosetrans} is indeed a solution of the system
of second order differential equations \eqref{wavebpsqk} with
a fixed value $\kappa=-1$ for $\ell=2$, and of a system
of first order equations for $\ell>2$. 

Thus the problem of determining the radial wave function of BPS black
holes is reduced to that of finding the appropriate class in 
$H^1(\cZ, \cO(-\ell))$. For a black hole with fixed electric and magnetic
charges $q_I, p_I$ and zero NUT charge, irrespective of $\ell$,
the only eigenmode of 
the generators \eqref{cxkillingh} is up to normalization
the ``coherent state''
\begin{equation}
\label{cohst}
g_{p,q}(\xi^I, \txi_I, \alpha) =  e^{i( p^I \txi_I
-q_I \xi^I)}\,.
\end{equation}
Applying the Penrose transform \eqref{penrosetrans} to the state
\eqref{cohst} using \eqref{gentwi}, we find (now labeling the different components
of the wavefunction by $\delta$)
\begin{equation}
\Psi^{(\delta)}_{p,q}(U, z^i, \bar z^{\bar j},\zeta^I, \tzeta_I, \sigma) = 
e^{i p^I \tzeta_I - i q_I \zeta^I} 
2^\ell\, e^{\ell U} \oint \frac{d z}{z} z^{\frac{\delta}{2}}
\exp \left[ e^{U} (z \bar{Z} + z^{-1} Z) \right],
\end{equation}
where $Z=Z_{p,q}(z^i,{\bar z}^{\bar j})$ 
is the central charge \eqref{centca}
\be
Z= e^{\cK(X,\bar X)/2} ( p^I F_I(X) - q_I X^I )\, ,
\ee
of the black hole.
After analytic continuation of $(\zeta^I,\tzeta_I)$ to 
$i(\zeta^I,\tzeta_I)$ and $(p^I,q_I)$ 
to $-i(p^I,q_I)$, as appropriate to the timelike reduction,
the integral may be evaluated in terms of a Bessel function,
\begin{equation}
\label{exactbpswf}
\Psi^{(\delta)}_{p,q}(U, z^i, \bar z^{\bar i}, \zeta^I, \tzeta_I, \sigma) 
= 2^{\ell+1}\pi\, e^{i p^I \tzeta_I - i q_I \zeta^I}
\ e^{\ell U}\ 
\left(\frac{\bar Z}{Z}\right)^{\frac{\delta}{4}}
\, J_{\frac{\delta}{2}}\left(2  e^U \abs{Z}\right)\, .
\end{equation}
This is the {\it exact} radial wave function for a black hole with 
fixed charges $(p^I,q_I)$, at least in the supergravity 
approximation\footnote{In the presence of $R^2$-type corrections,
the geodesic motion receives higher-derivative corrections, and
it is no longer clear how to quantize it.}.

Before analyzing the physical content of \eqref{exactbpswf},
it is worthwhile pointing out that it agrees in the 
semi-classical limit with direct quantization of 
the attractor flow equations \eqref{att}:
Identifying $dU/d\tau = p_U/2$ and $dz^i/d\tau=g^{i\bar j} p_{\bar j}$,
and quantizing the canonical momenta $p_U$ and $p_{\bar j}$ as
derivative operators $\frac{1}{i}\pa_U$ and 
$\frac{1}{i}\pa_{{\bar z}^{\bar j}}$
acting on $\Psi_{p,q}(U,z^i,{\bar z}^{\bar j})$, \eqref{att}
becomes
\be
\label{qatt}
\pa_U \Psi_{p,q} = e^{U} \, |Z| \, \Psi_{p,q}\ ,\quad
\pa_{{\bar z}^{\bar j}} \Psi_{p,q} = 
2 ~e^U~ \pa_{\overline \jmath} |Z| \, \Psi_{p,q}\ ,
\ee
which integrates to $\Psi_{p,q}=\exp( 2 i e^{U} |Z|)$. In the limit 
$U\to \infty$,  the phase of the wave function is stationary at the classical
attractor point (or points, should there be different basins of attraction),
as expected.

In the opposite near-horizon limit $U\to -\infty$, the
effective Planck constant $\hbar \sim e^{-U}$ goes to infinity,
leading to large quantum fluctuations. The exact result \eqref{exactbpswf} 
for the wave function is well behaved at the horizon,
\be
\Psi_{p,q}( U\to -\infty) \sim 
e^{(\ell-\delta/2) U}\,Z_{p,q}^{-\delta/2}/\Gamma(1+|\delta|/2)\ ,
\ee
but for fixed $U$ is not peaked at the attractor values of the flows. Instead,
it has local extrema whenever $|Z|$ does. This behavior may
seem at odds with the classical attractor behavior. 
The resolution of this paradox
is that the radial evolution of the moduli corresponds to 
the motion in an {\it inverted} potential $V=-e^{2U}V_{BH}$, 
which flattens out in the near horizon limit $U\to-\infty$.
(In fact, the radial flow is attractive in the 
BPS sector because it reduces to a gradient flow. In the
non-BPS sector, it is only attractive for extremal black
holes, at the cost of an infinite fine tuning of the initial
velocities~\cite{Tripathy:2005qp}.)

By reintroducing the cone variable $r$, it is also possible to study
the fluctuations of the horizon area. Setting $C^2=0$ in  \eqref{sepr}, 
the complete wave function on $\IR^+\times \cM_3$ is
\be
\Psi^{(\delta)}_{p,q}(r,U,z_i,{\bar z}_{\bar j}) \sim \ e^{\ell U}\,\left(\frac{\bar Z}{Z}\right)^{\frac{\delta}{4}}
\,J_{\frac12}(r)\,J_{\frac{\delta}{2}}\left(2  e^U \abs{Z}\right)\, .
\ee
Setting $A=4\pi r^2 e^{-2U}$, this may be translated into a wave function
for the moduli $z_i$ and the area,
\be
\Psi^{(\delta)}_{p,q}(r,U,z_i,{\bar z}_{\bar j}) \sim \ e^{\ell U}\,\left(\frac{\bar Z}{Z}\right)^{\frac{\delta}{4}}\,J_{\frac12}\left(
e^U \sqrt{A/4\pi} \right)\,
J_{\frac{\delta}{2}}\left(2  e^U \abs{Z}\right)\, .
\ee
In the limit $U\to +\infty$, the phase is stationary with respect to 
$U$ at $A/4\pi=\pm |Z|^2$ in agreement with classical expectations. 
At fixed $U$ however, the wave function is factorized and maximal around
$A=0$.

At this stage, we can now discuss the norm of the wave function. 
Under the Penrose transform, the Klein-Gordon
inner product on $\cM_3$  may be rewritten in terms of the 
holomorphic function $g$ as
\begin{equation}
\label{inner}
\inprod{\Psi \vert \Psi'} = \int 
d\xi^I d\txi_I d\alpha \, d\bar{\xi}^I d\bar{\txi}_I 
d\bar{\alpha}\ e^{(\ell-2n_V-4) K_\cZ}\,\overline{g(\xi^I, \txi_I, \alpha)}
\, g'(\xi^I, \txi_I, \alpha)\, ,
\end{equation}
where the integral runs over values of $\xi^I, \txi_I, \alpha,
\bar\xi^I, \bar\txi_I, \bar\alpha$ such that the bracket in 
\eqref{kz-hesse} is strictly positive. Moreover 
quantization of the electric, magnetic and NUT charges implies that
the integral over the real parts of $\xi^I, \txi_I, \alpha$ should run
over a fundamental domain of the Heisenberg group.
As announced at the end of Section
\ref{sphgeoq}, the inner product \eqref{inner} is formally positive 
definite\footnote{Equation \eqref{inner} 
is only formal, because $g$ and $g'$ are not well defined functions but 
rather representatives for cohomology classes.  To make it well defined, 
the integration region in \eqref{inner} has to be analytically continued
and interpreted in terms of contour integrals, but after doing so it is
not obviously positive definite anymore. For symmetric spaces, the unitarity
of the corresponding representations has been proven 
in some cases \cite{MR1038279,MR1421947}.}. 

While we have not proven normalizability of the exact wave function
\eqref{cohst}, its norm (if finite) 
is clearly unrelated to the exponential of the entropy:
Choosing for $g$ and $g'$ two coherent 
states of charges $(p,q)$ and $(p',q')$ as in \eqref{cohst}, 
the integral over the real parts of $(\xi,\txi,\alpha)$ gives a 
product of Kronecker deltas  $\delta_{p,p'}\delta_{q,q'}$ so
the remaining integral is
\be
\int d\zeta\, d\tzeta\,d\sigma\,
\left[ \Sigma^2( \zeta^I, \tzeta_I) + \frac{1}{16}
\sigma^2 \right]^{\frac{\ell}{2}-(n_V+2)}\,
\exp\left( - p^I \tzeta_I - q_I \zeta^I \right)\, ,
\ee
($\zeta^I,\tzeta_I,\sigma$ now represent the imaginary
parts of $\xi^I,\txi_I,\alpha$). For generic values of $p,q$, this
integral converges at infinity, while it converges at the origin
for large enough $\ell$. Homogeneity guarantees that the final
result, if finite, will be a homogeneous function  of the charges $p,q$, of 
degree $2\ell-2(n_V+2)$.

\section{Discussion \label{sec-discuss}}
In this paper, we have laid out a systematic framework for the radial,
mini-superspace, quantization of stationary, spherically symmetric
four-dimensional black holes.
 The key device was the equivalence
between radial evolution equations and geodesic motion on the moduli
space after reduction along three-dimensions. This equivalence holds
in general for gravity theories with an arbitrary number of Maxwell
fields and scalar fields at two-derivative order, and does not assume 
any supersymmetry. It offers a direct path towards quantization, subject
to the usual canonical quantum gravity caveats. It is worth stressing that
the wave function of a generic black hole with fixed charges is 
by no means unique, nor should it be.

In the context of $\cN=2$ supergravity, we have shown that the 
phase space of BPS solutions is isomorphic to the twistor space $\cZ$
of the moduli space~$\cM_3$ of the three-dimensional theory, making it manifest
that the BPS constraints are first class. We have proposed to
identify the BPS Hilbert space as the K\"ahler quantization of 
$\cZ$, with the necessary amendments due to the non-positive 
definiteness of the metric on $\cZ$. This proposal is mathematically
natural in view of the Penrose transform, which relates cohomology
classes valued in a line bundle $\cO(-\ell)$ on $\cZ$ to
solutions of a set of linear partial differential equations 
on  $\cM_3$, which agree with the BPS constraints
in the semi-classical limit (at least when $\ell=2$). Ordering ambiguities are not entirely
resolved, but parameterized by the undetermined integer
parameter $\ell$ corresponding to the spin of the wave function
under the $SU(2)_R$ symmetry group, which could in principle be
determined by a more careful treatment of the one-dimensional 
non-linear sigma model on $\cM_3$ including fermions and ghosts (see~\cite{w-toappear}).  
In this framework, the wave function is uniquely determined, 
and agrees with semi-classical expectations in the limit 
far from the horizon. In the near-horizon limit, quantum fluctuations
become dominant, as the effective Planck constant is given by 
$\hbar\sim e^{-U}$.

This systematic study enables us to examine the suggestion in 
\cite{Ooguri:2005vr} to identify the topological string amplitude
as the radial wave function for BPS black holes. Taken literally, 
this statement cannot be true in our framework,
if only because the functional
dimension of the Hilbert spaces of the BPS radial quantization
($2n_V+3$) and of the topological amplitude ($n_V+1$) are so
different. Moreover, the electric and magnetic charge operators
of BPS black holes can be simultaneously diagonalized (when the
NUT charge vanishes), whereas the corresponding operators in the
topological Hilbert space are inherently non-commutative.
One may however try to rescue the suggestion in \cite{Ooguri:2005vr}
by noting that, after lifting the geodesic motion on 
$\cM_3^*$ to the Swann space $\cS$ (which includes the Killing spinor
$\eps^{A'}$ on top of the usual moduli), there exists 
an even smaller subspace of the general phase space $T^*(\cS)$,
namely the  $2n_V+4$-real dimensional subset of the Swann space 
where the anti-holomorphic
coordinates $\bar z^{\bar \aleph}$ take a fixed (arbitrary)
value. Since $\cS$ is hyperk\"ahler, it is in particular holomorphic
symplectic, and the above mentioned space has a natural symplectic
form. Its quantization would in principle lead to a ``super-BPS'' Hilbert space
of functional dimension $n_V+2$, just one over the dimension $n_V+1$
of the topological Hilbert space. Geometrically, it should be defined
by a kind of ``tri-holomorphy'' condition on $\cS$ (just as the 
regular BPS Hilbert space corresponds to holomorphic functions, or
sections, on $\cZ$) whose precise definition is left to future work.
If correct, this proposal leads to a one-parameter generalization
of the topological string, first outlined in \cite{Gunaydin:2006bz},
describing F-term couplings in $\cN=4$ supergravity 
on the vector and hypermultiplet branches in three dimensions.
The extra parameter can be thought of as the NUT charge $k$, the
scale $U$ of the thermal circle, or, in the T-dual picture, as the
string coupling in four dimensions.

The framework discussed in this paper is quite general,
and can be further extended in many different directions,
some of which we hope to address in future publications:

\begin{enumerate}
\item Some of the considerations above can be made more explicit
in a special class of $\CN=2$ supergravities with symmetric moduli
spaces, ${\cal M}_4=G_4/U(1)\times H$ and ${\cal M}_3=G_3/SU(2)\times M$.  
This happens when the prepotential $F$ is equal to the cubic norm 
of a Euclidean Jordan algebra $J$ of degree three, in which case the four-dimensional 
U-duality group $G_4$ is simply the conformal group of $J$ that acts by analytic automorphisms on the Hermitian symmetric space ${\cal M}_4=G_4/U(1)\times H$
and leaves a light cone defined by the cubic norm invariant \cite{Gunaydin:1983rk,Gunaydin:1983bi,Gunaydin:1984ak,Ferrara:1997uz}.
The corresponding three-dimensional duality group  $G_3$ is of quaternionic noncompact real form \cite{Gunaydin:1983rk,Gunaydin:1983bi} and  
can be constructed as the invariance group of a "light-cone" defined by a quartic 
norm associated with $J$ \footnote{ More precisely, the quartic form is defined over the Freudenthal system defined by $J$.}~\cite{Gunaydin:2000xr}. 
Some of these very special supergravity theories
are known to correspond to the low energy limit of string theories, 
such as the FHSV model with $G_4=Sl(2,\IR)\times SO(2,10)$,
$G_3=SO(4,12)$ \cite{Ferrara:1995yx}. In such cases, the BPS
and ``super-BPS'' Hilbert spaces furnish a special family of
unitary representations of $G_3$, which
as we explain in a separate paper~\cite{Gunaydin:2007qq}, correspond
to the ``quasi-conformal'' and ``minimal'' representations 
previously constructed in the literature.
$G_3$ being a solution-generating symmetry for black holes in four dimensions,
it is natural to assume that a discrete subgroup $G_3(\IZ)$
remains as a spectrum-generating symmetry in the putative quantum
theory reducing to this very special supergravity at 
low energies \cite{Ferrara:1997uz,Gunaydin:2000xr,Pioline:2005vi,
Gunaydin:2005mx}. This suggests that the partition function
for the exact BPS black hole degeneracies should be an automorphic
form of $G_3(\IZ)$, attached to the above unitary representation.

\item The same strategy can be applied to BPS black holes
in $\CN>2$ supergravities, where the moduli spaces in 4
and 3 dimensions are always symmetric. 
The details however differ, since the relevant twistor
spaces are no longer two-sphere bundles, and the duality
groups $G_4$ and $G_3$ are in different real forms (e.g.
the split real form for $\cN=8$). Moreover, there will exist
different BPS Hilbert spaces depending on the number of supersymmetries
left unbroken by the black hole (In Appendix \ref{ngt2}, we sketch 
some basic features of these constructions). It would be 
interesting to understand in more detail the corresponding 
unipotent representations of $G_3$, construct explicit automorphic forms
attached to these representations and compare their Fourier
coefficients with the microscopic degeneracies. This would generalize and 
possibly amend the approach in~\cite{Dijkgraaf:1996it} for 1/4-BPS dyons 
in $\cN=4$ string theory, opening the possibility to switch on 
chemical potentials for each electric or magnetic charge separately. 

\item It is also of interest to apply this framework to non-BPS, 
extremal black holes, corresponding 
to more general light-like geodesics on $\cM_3^*$,
not satisfying the holomorphy conditions. In view of their 
attractor behavior, it may be interesting
to investigate whether their wave function still exhibits some
universality properties as $U\to \infty$. Black holes in 
gauged supergravities would also be interesting to analyze.

\item Multi-centered black holes are more challenging. Assuming stationarity,
the reduction to the non-linear sigma model on the three-dimensional
moduli space still goes through. It would 
be interesting to formulate the general multi-centered solutions 
as holomorphic maps from $\IR^3$ to the Swann or twistor space over
$\cM_3$, and possibly generate new solutions in this fashion. It is
also reasonable to expect that there should be a ``multi-particle''
picture for multi-centered black holes, in terms of forked geodesics
on $\cM_3^*$. The quantization of these BPS solutions would then 
amount to the ``second quantization'' of the one-black hole
BPS Hilbert space.

\item By T-duality along the thermal circle, the \qk moduli space
arising from the reduction of the vector multiplets to three
dimensions is related to the hypermultiplet moduli space 
in the dual string theory in four dimensions. It would be interesting
to relate the black hole wave function to D-instanton contributions to
couplings on the hypermultiplet branch satisfying the same generalized
harmonicity conditions~\cite{npv-to-appear}.

\end{enumerate}

\subsection*{Acknowledgements}

It is our pleasure to thank S.~Minwalla, M.~Ro$\check{\rm c}$ek
and
especially S.~Vandoren for useful discussions.  M.G. and B.P. express
their gratitute to the organizers of the program ``Mathematical Structures
in String Theory'' that took place at KITP in the Fall of 2005, where
this work was initiated.
A.W. thanks LPTHE Jussieu for warm hospitality.
B.P.'s research
is supported in part by  the EU under contracts
MTRN--CT--2004--005104, MTRN--CT--2004--512194, and by
ANR (CNRS--USAR) contract No 05--BLAN--0079--01.
The research of
A.N. is supported by the Martin A. and Helen Chooljian Membership at the
Institute for Advanced Study and by NSF grant PHY-0503584.
The research of M.G.
was supported in part by the National Science Foundation under grant number PHY-0555605 and  support of Monell Foundation  during his sabbatical stay at IAS, Princeton, is gratefully acknowledged. Any opinions, 
findings and conclusions or recommendations expressed in this material are those of the authors and do not 
necessarily reflect the views of the National Science Foundation.

\appendix

\section{Reducing the Supersymmetry Conditions \label{redferm}}
In this Appendix, we discuss the dimensional 
reduction of the supersymmetry conditions
in four dimensional, $\cN=2$ supergravity on the time-independent 
ansatz \eqref{stationan}, and further on 
the spherically-symmetric ansatz \eqref{3dsli}.

The supersymmetry transformations of
the four dimensional gravitini and gaugini $(\psi_\mu, \lambda^a)$
to leading order in fermi fields are
\bea
\delta \psi_{R\mu} &=&\ \quad {\cal D}_\mu \varepsilon_R \ \
+ \ \ \frac14 \ e^{K/2} X^I(\Im{\cal N})_{IJ} F^J_{\nu \rho} \gamma
^{\nu\rho}\gamma_\mu\varepsilon_L\, ,\nn\\[2mm]
\delta \lambda^a_R\  &=& -\frac12 e^a{}_i \gamma^\mu \partial_\mu z^i\varepsilon_L + \frac 14 \  \bar{f}^{aI} 
(\Im{\cal N})_{IJ }F^J_{\nu\rho}\gamma^{\nu\rho}\varepsilon_R\, .
\label{variations}
\eea
Here,  the gravitino $\psi_\mu$, gaugini $\lambda^a$ and
supersymmetry parameter $\varepsilon$ are four-dimensional complex Dirac
spinors, and the subscripts $L,R$ denote their
chiral projections under $L,R=\frac12(1\pm \gamma^5)$. 
The derivative ${\cal D}=dx^\mu {\cal D}_\mu=D+Q$ is the sum of the 
Levi-Civita connection $D$ and K\"ahler connection $Q$, with
\be
Q=\frac 14 (\partial - \bar \partial)K=-\frac 14  
\frac{\bar X N dX -  \overline{dX} N X}{\bar X N X}\, ,
\ee 
and $N_{IJ}\equiv (\Im\tau)_{IJ}$. 
Solutions preserve some amount of supersymmetry when there
exists a non-zero ``Killing spinor'' $\varepsilon$ such that the right-hand 
sides of~\eqref{variations} vanish.

To reduce the four dimensional variations~\eqref{variations} to three, 
and in turn one dimension,
we begin by collecting some useful data:
The spin connection and Dirac matrices in the timelike reduction ansatz~\eqref{stationan}
are
\bea
\gamma^t=e^{-U}\gamma^0-e^U\omega_{\ri}\, {}^{(3)\!}\gamma^\ri\, ,&\quad&
\gamma^\ri= e^U\, {}^{(3)\!}\gamma^\ri\, ,\nn\\[2mm]
\omega_{0\rb}=-e^U e^0 \partial_\rb U + \frac 12 e^U dx^\ri F_{\ri \rb}\, ,&&
\omega_{\ra\rb}={}^{(3)\!}\omega_{\ra\rb}+\frac12 e^{3U} e^0 F_{\ra\rb} -2 
dx_{[\ra}\partial_{\rb]} U\, .
\eea
Here four dimensional curved and flat indices decompose as 
$\mu,\nu,\ldots=(t;\ri,\rj...)$, 
$m,n,\ldots$ $=$ $(0;\ra,\rb\ldots)$ while 
$F_{\ri\rj}=2\partial_{[\ri}\omega_{\rj]}$ is the graviphoton field strength 
and $e^m=(e^0,e^U{}^{(3)\!}e^\ra)$ with $e^0=e^U[dt+\omega]$  the timelike
vierbein. All three dimensional indices are manipulated with the three 
dimensional metric and 
drei-bein. For dualizing we use the identity
\be
\gamma^{\ri\rj}=\frac{ie^{2U}}{\sqrt{{}^{(3)\!}g}}\varepsilon^{\ri\rj\rk}
\ {}^{(3)\!}\gamma_\rk\gamma^0\gamma^5\, ,
\ee
plus the relations between magnetic field strengths and magnetic potentials
\bea
 F\  \equiv \ \star_3 d\omega\,  &=&\ \ -e^{-4U}(d\sigma +\zeta^I d\zetat_I-\zetat_I d\zeta^I)\, ,\nn\\[2mm]
 F^I \equiv\star_3 d A^I_3&=&-\zeta^I F + e^{-2U}(\Im{\cal N})^{IJ}\ (d\zetat_J+(\Re {\cal N})_{JK}
d\zeta^K)\, .
\eea
Equipped with the above data, the reduced spinor-covariant derivative 
is easily computed
\be
dx^\mu D_\mu\equiv D ={}^{(3)}\!D+
\frac12 e^{U}(e^0 \, {}^{(3)\!}\gamma^{\rj}\gamma^0-e^{-U}dx_\ri\, {}^{(3)\!}\gamma^{\ri\rj} )
(\partial_\rj U - \frac i2 \gamma^5 e^{2U} F_\rj)\, .
\ee
It also pays to calculate
\be
F^I_{\nu\rho}\gamma^{\nu\rho}
=2\, {}^{(3)\!}\gamma^\ri\gamma^0(\partial_\ri \zeta^I +i \gamma^5 
e^{2U}[F^I_\ri+\zeta^I F_\ri]
)\, .
\ee
Then orchestrating the gravitini variations parallel to the timelike
 vierbein $e^0$ along with the gaugini variations, we find
\bea
0&=&
i  e^{K/2-U} X^I
[\slashed{\partial}\zetat_I+{\cal N}_{IJ}\slashed{\partial}\zeta^J] \varepsilon_L- [
\slashed{\partial} U - \frac i2 e^{-2U} (\slashed{\partial}\sigma -\zeta^I \slashed{\partial}
\zetat_I-\zetat_I \slashed{\partial}\zeta^I)]  \gamma^0 \varepsilon_R
\, ,\nn\\[5mm]
0&=&\quad  e^a{}_i\slashed{\partial} z^i \varepsilon_L \ \ + \ \ i e^{-U} \bar f^{aI}
[\slashed{\partial}\zetat_I+{\cal N}_{IJ}\slashed{\partial}\zeta^J]
\gamma^0 \varepsilon_R\, ,
\eea
with $\slashed\partial\equiv {}^{(3)\!}\gamma^\ri\partial_\ri$ the 
three dimensional Dirac operator. 
Comparing to the expressions for the quaternionic vielbein in~\eqref{vb} 
yields the three dimensional Killing spinor equations
\be
0=
\begin{pmatrix}i \slashed u  & &  \slashed v \\
 \slashed e^a & &i \slashed E^a  \\
-i\slashed {\bar E}^{\bar a} & & \slashed{\bar e}^{\bar a} 
\\ -\slashed{\bar v} & &i \slashed{\bar u}\end{pmatrix} 
\left(
\begin{array}{c}
\varepsilon_L\\
\gamma^0\varepsilon_R
\end{array}\right) = {}^{(3)\!}\gamma^\ri V_\ri^{AA'}\varepsilon_{A'}\, ,
\ee
where the one-forms $V_a^{AA'}$ on $\cM_3$ have been pulled back to the spatial
slice. This result is consistent with the  SUSY transformations 
in~\eqref{dpsi}.

We must still examine terms proportional to $dx^{\ri}$ in the 
gravitini variations.
Terms involving antisymmetrized pairs of Dirac matrices do not yield 
independent equations, so we find
\be
0=\left[{}^{(3)\!}{\cal D}
+
\left(
\begin{array}{cc}
-\frac12 \bar v&i \bar u\\
\!-iu \ &-\frac12 v
\end{array}\right)
\right]
\left(
\begin{array}{c}
\varepsilon_L\\
\gamma^0\varepsilon_R
\end{array}\right)\, .
\ee
Defining rescaled SUSY parameters $\epsilon$
\be
\varepsilon = \left(
\begin{array}{c}
\varepsilon_L\\
\gamma^0\varepsilon_R
\end{array}\right)
\equiv e^{-U/2}\left(\begin{array}{c}
\epsilon^1\\
\epsilon^2
\end{array}\right)=e^{-U/2}(\epsilon^{A'})\, ,
\ee
yields
\be
0=\left[{}^{(3)\!}D
+
\left(
\begin{array}{cc}
\frac14 (v-\bar v)+Q&i \bar u\\
\!-iu \ &-\frac14 (v-\bar v)-Q
\end{array}\right)
\right]
\left(
\begin{array}{c}
\epsilon^1\\
\epsilon^2
\end{array}\right)
\equiv \Big({}^{(3)\!}D \epsilon^{A'} + p^{A'}_{B'} \epsilon^{B'}\Big)\, ,
\label{DE}
\ee
where $p^{A'}_{B'}$ is the $sl(2,{\mathbb R})$ 
valued connection
over the ${\cal M}^*_3$ moduli space. 
Indeed the Swann bundle is obtained as a ${\mathbb C}^2$ fibration with this connection
over ${\cal M}^*_3$. But first we need to compute the reduction from three dimensions
to one quantum mechanical dimension. Let us pause to collect the Killing spinor 
equations in three dimensions:
\be
{}^{(3)\!}\gamma^\ri V_i^{AA'}\epsilon_{A'}=0={}^{(3)\!}D \epsilon^{A'} 
+p^{A'}_{B'} \epsilon^{B'}\, .\label{pause}
\ee

The $3\rightarrow 1$ reduction proceeds along the ansatz~\eqref{3dsli}
whose dreibeine and spin connections are
\bea
e^1=Nd\rho\, ,\quad e^2=rd\theta\, ,\quad e^3=r\sin\theta d\varphi\, ,\quad\nn\\[3mm]
\omega^{12}=\frac{r'd\theta}{N}\, , \quad 
\omega^{13}=\frac{r'\sin\theta d\varphi}{N}\, , \quad
\omega^{23}=-\cos\theta d\varphi\, .\eea
We compute the covariant exterior derivative acting on spinors:
\be
{}^{(3)}\! D={}^{(2)} D + d\rho \frac\partial{\partial\rho}-\frac i2 \frac{r'}{N}\ {}^{(2)}\!\sigma\, .
\ee
Here the covariant exterior derivative on the sphere is
\be
{}^{(2)} D =d\theta\frac\partial{\partial\theta}+d\varphi\frac\partial{\partial\varphi}-
\frac 12 \sigma^1\sigma^2 \cos\theta d\varphi\, ,
\ee
where the two dimensional Dirac matrices are $\sigma^1=i\gamma^1\gamma^2$,
$\sigma^2=i\gamma^1\gamma^3$, 
and 
$
{}^{(2)}\!\sigma=\sigma^1 \ {}^{(2)}\!e^1+\sigma^2 \ {}^{(2)}\!e^2=
\sigma^1 d\theta + \sigma^2\sin\theta d\varphi\, . 
$
We make the ansatz
\be
\epsilon^{A'} = \pi^{A'}(\rho) \chi\, ,\qquad
\ee
where $\chi$ is a vector in the two dimensional space of 
complex Killing spinors on $S^2$, obeying
\be
{}^{(2)}\!D \chi = \frac12 {}^{(2)}\! \sigma \chi\, .
\ee
and all other fields $(N,r,\zeta^I,\tilde \zeta_I,a,U)$ depend only on $\rho$.
This allows us to split~\eqref{DE} into its radial and spherical parts. Requiring
that $\chi$ be the most arbitrary Killing spinor on the sphere
the three dimensional Killing spinor equations~\eqref{pause} reduce to
\bea
0&=&d\rho \ V_\rho^{AA'} \pi_{A'}\, ,\nn\\[2mm]
0&=&d\pi^{A'} + p^{A'}_{B'} \pi^{B'}\, ,\nn\\[2mm]
0&=&\frac{dr}{N}-d\rho\, .
\eea
reproducing \eqref{bpscondhk} and \eqref{rnsusy}.

\section{Minimal $\cN=2$ Supergravity}
\begin{figure}
\centerline{
\hfill\includegraphics[height=5cm]{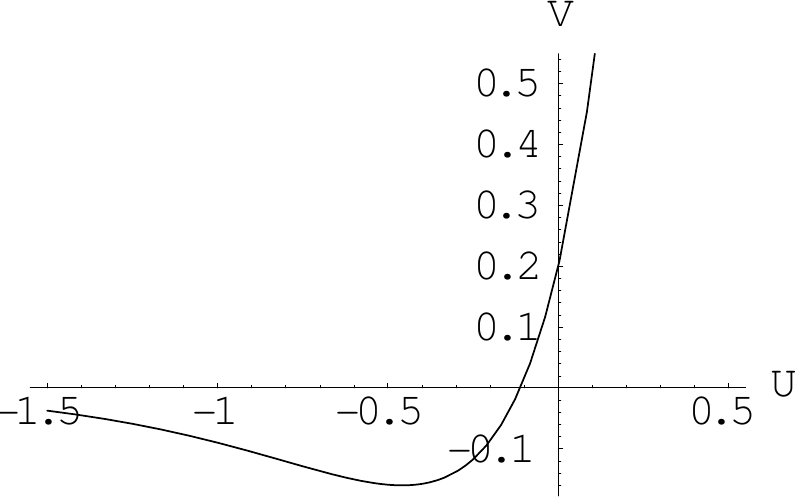}
\hfill\includegraphics[height=5cm]{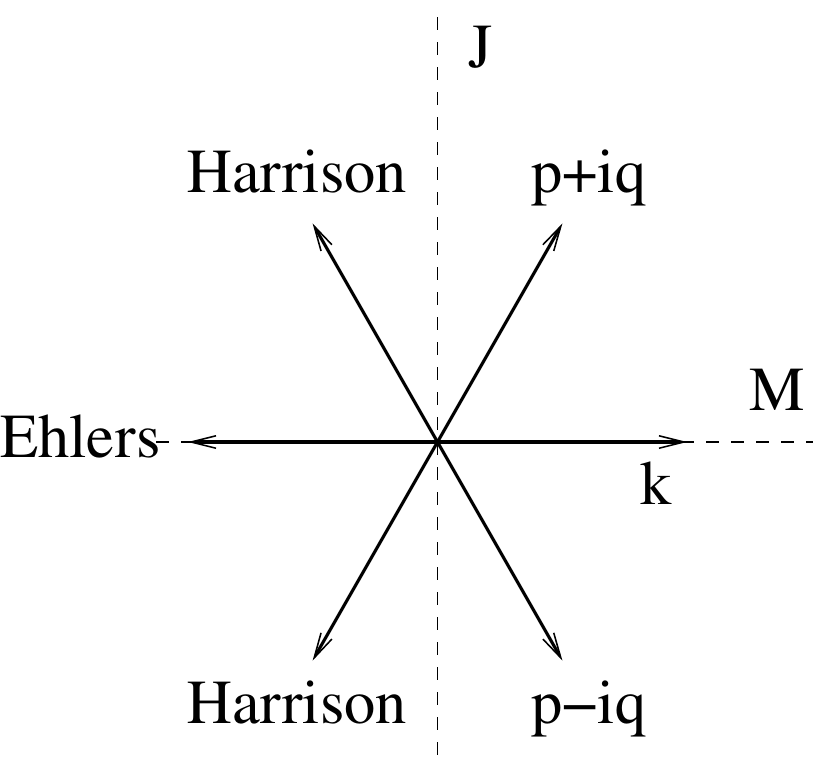}
\hfill}
\caption{{\it Left:} Potential governing the motion along
the $U$ variable in the universal sector. The horizon
is reached at  $U\to-\infty$. {\it Right}: Root
diagram of the $SU(2,1)$ symmetries in the universal sector.\label{su21pic}}
\end{figure}

In this Appendix, we work out the details for minimal $\cN=2$ supergravity
in four dimensions, with no vector multiplet, and trivial 
prepotential $F=-i (X^0)^2$. The resulting moduli space 
in three dimensions is the symmetric space
${\cal M}_3^*=SU(2,1)/Sl(2)\times U(1)$,
or its analytic continuation of the quaternionic-K\"ahler space
${\cal M}_3=SU(2,1)/$ $SU(2)\times U(1)$.
The same $\cM_3$ describes the tree-level couplings of the universal 
hypermultiplet in 4 dimensions.
The classical Hamiltonian \eqref{lapbel} reduces to
\be
\label{hamuni}
H =  \frac14(p_U)^2 - \frac12 e^{2U} 
\left[ ( p_{\tilde\zeta} -  k \zeta )^2 
+ (p_\zeta + k \tilde\zeta)^2 \right]
+ \frac14 e^{4U} k^2\ .
\ee
The motion separates between the $(\tzeta,\zeta)$ plane and the $U$ direction, 
while the
NUT potential $\sigma$ can be eliminated in favor of its conjugate momentum
$k=2e^{-4U}(\dot \sigma +  \tzeta \dot\zeta - \zeta \dot\tzeta)$.
The potential is depicted on Figure \ref{su21pic} (left).
The motion in the $(\tzeta,\zeta)$ plane is that of a 
charged particle in a constant magnetic field. The electric, magnetic charges
and the angular momentum $J$ in the plane (not to be confused with
that of the black hole, which vanishes by spherical symmetry)
\be
p = p_{\tilde \zeta} +  \zeta k \ ,\quad q = p_{\zeta} -  \tilde\zeta k
\ ,\quad J = \zeta p_{\tzeta} - \tzeta p_\zeta\, ,
\ee
satisfy the usual algebra of the Landau problem,
\be
[p,q] = -2k\, ,\quad [J,p]=q\ ,\quad [J,q]=-p\, ,
\ee
where $p$ and $q$ are the ``magnetic translations''.
The motion in the $U$ direction is governed effectively by
\be
H =  \frac14 (p_U)^2 + \frac14 e^{4U} k^2
- \frac12 e^{2U} \left[ p^2 + q^2 -4 k J\right] = C^2\, .
\ee
At spatial infinity ($\tau=0$), one may impose the initial conditions
$U=\zeta=\tzeta=\sigma=0$. The momentum $p_U$ at infinity equals the 
ADM mass, and $J$ vanishes, so the mass shell condition becomes
\be
\label{bpsuh}
\frac14 m^2 + \frac14 k^2 - \frac12(p^2 + q^2) = C^2\, .
\ee
In this simple case, the extremality condition $C^2=0$ is equivalent
to supersymmetry, since the vielbein $V$ is a $2\times 2$ matrix.
Equation~\eqref{bpsuh} is the BPS mass condition, generalized
to non-zero NUT charge. Note that for a given value of $p,q$,
there is a maximal value of $k$ such that $M^2$ remains positive.

At the horizon $U\to -\infty$, $\tau\to\infty$, the last term 
in~\eqref{hamuni} is irrelevant, and one may integrate the
equation of motion of $U$, and verify that the metric~\eqref{stationan}
becomes $AdS_2\times S^2$ with area
\be
A =  4\pi (p^2 + q^2)\, ,
\ee
recovering the usual entropy $S=A/4$ of Reissner-Nordstr\"om
black holes.

Since the universal sector is a symmetric space, there must exist
three additional conserved charges, so that the total set of conserved
charges can be arranged in an element $Q$ in the Lie algebra 
${\mathfrak g}_3=su(2,1)$ 
(or rather, in its dual ${\mathfrak g}_3^*$). The physical origin of
these are the Ehlers and Harrison transformations \cite{kinnersley}.
The root diagram
of $SU(2,1)$ is depicted on Figure \ref{su21pic}. 
The Casimir invariants of~$Q$ are easily computed from the explicit
form of the Killing vectors~\cite{Gunaydin:2007qq}:
\be
\Tr (Q^2)= H \ ,\quad \det(Q)=0\, .
\ee
The last condition ensures that the conserved quantities do not 
overdetermine the motion. 
The co-adjoint action $Q\to h Q h^{-1}$ of $G_3$ on ${\mathfrak g}_3^*$
relates different trajectories with the same value of $H$.
The phase space, at fixed value of $H$, is therefore the
co-adjoint orbit of a diagonalizable element of  ${\mathfrak g}_3^*$, 
of dimension 6 (the
symplectic quotient of the full 8-dimensional phase space
by the Hamiltonian $H$).
By the Kirillov-Kostant construction, it carries a
canonical symplectic form such that the Noether charges
represent the Lie algebra ${\mathfrak g}_3$. 

As we have just seen, BPS solutions have $H=0$. The Cayley--Hamilton property
for $3\times 3$ matrices
\be
Q^3 - \Tr(Q) Q^2 - \frac12[\Tr(Q^2)-(\Tr Q)^2] Q - \det(Q) = 0\, ,
\ee
then implies that $Q^3=0$ as a matrix equation in the fundamental 
representation. $Q$ is therefore non-diagonalizable, with Jordan
normal form 
\be
\begin{matrix}
Q = h\cdot \begin{pmatrix} 0&1&0 \\ 0 & 0 & 1 \\ 0 & 0 & 0 
\end{pmatrix}\cdot h^{-1}
\end{matrix}\, .
\ee
The stabilizer of the Jordan block is the parabolic group of lower triangular
unimodular matrices $P$. The BPS phase space is therefore $Sl(3,\IC)/P_{\IC}$,
which is indeed the twistor space 
of\footnote{It is a 
peculiarity of this model that the  BPS and generic phase spaces are both
six dimensional.}
$\cM_3$ .
Upon quantization, one finds that the BPS Hilbert space corresponds
to the quaternionic discrete series of $SU(2,1)$ \cite{gnppw-to-appear}.
The ``super-BPS'' phase space corresponds to the case where $Q$ is
nilpotent of degree 2 ($Q^2=0$), and leads to one of 
the minimal representations of $SU(2,1)$.

The pattern found here continues to hold in very special $\cN=2$ 
supergravities, where ${\cal M}_3=G_3/SU(2)\times M$ is a symmetric 
quaternionic-K\"ahler space. The BPS phase space is the twistor
space $\cZ=G_3/U(1)\times M$. The sheaf cohomology $H^1(\cO(-\ell),\cZ)$,
for $\ell$ large enough, furnishes a unitary representation of $G_3$ 
of functional dimension $2n_V+3$, belonging to the
quaternionic discrete series. For one special value of $\ell$, it
admits an irreducible submodule which furnishes the minimal
representation of $G_3$, of functional dimension $n_V+2$.

\section{BPS Black Holes and Geodesic Motion in  $\cN>2$ SUGRA \label{ngt2}}
In this Appendix, we confine ourselves to some preliminary remarks
about the extension of our formalism to supergravity theories with 
$\N\geq 2$ supersymmetry in 4 dimensions. A common feature, shared
with the $\cN=2$ very special supergravity theories, is that
the moduli spaces $\cM_3=G/K$ and $\cM_3^*=G/K^*$
are symmetric (resp. affine symmetric) spaces, and amenable to group
and representation theory methods. A generalization of the 
twistor space construction for non-quaternionic, symmetric spaces
has been studied in \cite{MR791300}. In general, there exist
different classes of BPS geodesics, depending on the number of
supersymmetries left unbroken by the black hole, and classified
by the orbit of the momentum $P$ under $K$.
By the Kostant-Sekiguchi correspondence \cite{MR867991,MR1328708}, 
orbits of $P$ under $K_{\IC}$ are in one-to-one correspondance with nilpotent
orbits of $G$ and in turn related to unitary representations of $G$ by
Kirillov's orbit philosophy \cite{MR1701415}. This provides a systematic
way to discuss the quantization of BPS black holes in  $\cN>2$ supergravity.

We start with $\CN=8$ supergravity in four dimensions.
The moduli space is the 70-dimensional 
symmetric space $\cM_4=E_{7(7)}/SU(8)$.
Upon reduction to three dimensions, either along a space-like or 
a time-like direction, one obtains the 128-dimensional spaces
\be
{\cal M}_3= \frac{E_{8(8)}}{SO(16)} \ ,\quad
{\cal M}_3^*= \frac{E_{8(8)}}{SO^*(16)}\, ,
\ee
where $SO^*(16)$ is the real form of $SO(16)$ with maximal non-compact
group $U(8)$.  The supersymmetry  variations of the
fermions in the non-linear sigma model on $\cM_3^*$ are~\cite{deWit:1992up}
\be
\delta\lambda_{A} = \eps_I \Gamma^I_{A \dot A} P^{\dot A} \, ,
\ee
where the SUSY parameter $\epsilon_I$ transforms in
a vector representation of the R-symmetry group $SO^*(16)$, 
the momentum $P^{\dot A}$ in a 128-dimensional 
real spinor representation of $SO^*(16)$ (corresponding to 
the tangent space to $E_{8(8)}/SO^*(16)$), and $\lambda_A$ is 
in the conjugate spinor representation $\overline{128}$.
Depending on the orbit of the momentum $P^{\dot A}$ under $SO^*(16)$,
the number of unbroken symmetries will be different. 
Half-BPS states, preserving 16 out of the 32 supersymmetries, 
are obtained when $P^{\dot A}$ is a pure spinor in Cartan's 
sense\footnote{Recall that
Cartan's pure spinor of $SO(2n)$ is isomorphic to $\IC \times SO(2n)/U(n)$.}.
This orbit has dimension $58$, and quantizes into the minimal representation
of $E_{8(8)}$ constructed in \cite{Kazhdan:2001nx,Gunaydin:2001bt},
with functional dimension 29. Quarter and 1/8-BPS black holes 
are associated to 92-dimensional and
114-dimensional orbits of spinors of lesser purity. In addition,
there is an 112-dimensional orbit corresponding to 1/8-BPS black holes
with zero entropy. Upon quantization, these reduced phase spaces
should lead to unipotent representations of $E_{8(8)}$ with functional dimensions
46, 57, and 56, respectively, which should be considered as the 
analytic continuations of corresponding representations of $E_{8(-24)}$
constructed in \cite{MR1421947}. It would be interesting
to lift the geodesic motion to the generalized twistor space $E_{8(8)}/SO(2)\times SO(14)$,
and determine in this way the most general 1/8-BPS black hole solution in 
four dimensions.

We now turn to $\CN=4$ supergravity with $n_v$ vector multiplets.
The moduli space in four dimensions is $SU(1,1)/U(1) \times SO(6,n_v)/SO(6)\times
SO(n_v)$. After compactification to three
dimensions, one obtains \cite{Breitenlohner:1987dg}\footnote{For $n_v=22$, 
the group $SO(8,n_v+2)$ is the spectrum-generating
symmetry that was used in~\cite{Cvetic:1995kv} to obtain the general
black hole solution in heterotic string  theory compactified on $T^6$.}
\be
{\cal M}_3 = \frac{SO(8,n_v+2)}{SO(8)\times SO(n_v+2)}\ ,\quad
{\cal M}_3^* = \frac{SO(8,n_v+2)}{SO(6,2)\times SO(2,n_v)}\, .
\ee
The supersymmetric variation of the fermions is now
\be
\delta\lambda_{A}^a = \epsilon_I \Gamma^I_{A \dot A} P^{\dot Aa} \, ,
\ee
where $\epsilon_I$ is a  vector of the
R-symmetry group $SO(6,2)$, and $P^{\dot Aa}$ ($a=1...n_v$), 
are a collection of $n_v$ spinors of $SO(6,2)$ corresponding to the
tangent space of $SO(8,n_v+2)/SO(6,2)\times SO(2,n_v)$. 
Supersymmetric solutions can be obtained by
requiring that the momentum factorizes into 
$P^{\dot Aa}  = \lambda^{\dot A} v^a$.
Half-BPS trajectories are obtained when
$\lambda^{\dot A}$ is a pure spinor of $SO(6,2)$, and
$v^a$ has zero norm. The complex dimension of the space of 
pure spinors of $SO(6,2)$ is $7$ while that of null vectors is $n_v+1$, 
so this orbit has complex dimension $n_v + 7$. Upon quantization,
we obtain the minimal representation of $SO(8,n_v+2)$, of real 
dimension $n_V+7$. It would be interesting to study the lift
of BPS geodesics to the generalized twistor space $SO(8,n_v+2)/[U(4)\times SO(n_v+2)]$.
Similar comments can be made for $\cN=3,6$ supergravity in four dimensions.

\bibliography{combined}
\bibliographystyle{utphys}

\end{document}